\documentclass[twocolumn,showpacs,preprintnumbers,amsmath,amssymb]{revtex4}
\usepackage{graphicx}
\usepackage{dcolumn}
\usepackage{bm}

\newcommand{\Slash}[1]{{\ooalign{\hfil/\hfil\crcr$#1$}}}

\begin{document}

\title{Scalar-Quark Systems and Chimera Hadrons in $\bf{SU(3)_c}$ Lattice QCD}
\author{H.~Iida$^1$\footnote{
E-mail: iida@yukawa.kyoto-u.ac.jp, iida@rcnp.osaka-u.ac.jp}, 
H.~Suganuma$^2$ and T.~T.~Takahashi$^1$}

\affiliation{$^1$
Yukawa Institute for Theoretical Physics, Kyoto University, 
Sakyo, Kyoto 606-8502, Japan}
\affiliation{
$^2$Department of Physics, Kyoto University, Graduate School of Science, 
Sakyo, Kyoto 606-8502, Japan}
\date{\today}
\begin{abstract}
In terms of mass generation in the strong interaction 
without chiral symmetry breaking, 
we perform the first study for light scalar-quarks $\phi$ 
(colored scalar particles with ${\bf 3}_c$ or idealized diquarks) 
and their color-singlet hadronic states using quenched SU(3)$_c$ lattice QCD 
with $\beta=5.70$ (i.e., $a\simeq 0.18{\rm fm}$) 
and lattice size $16^3\times 32$. 
We investigate ``scalar-quark mesons'' $\phi^\dagger \phi$ and 
``scalar-quark baryons'' $\phi\phi\phi$ as 
the bound states of scalar-quarks $\phi$.
We also investigate the color-singlet bound states of scalar-quarks 
$\phi$ and quarks $\psi$, i.e., 
$\phi^\dagger \psi$, $\psi\psi\phi$ and $\phi\phi\psi$, which we name 
``chimera hadrons''. 
All the new-type hadrons including $\phi$ are found to have a large mass 
even for zero bare scalar-quark mass $m_\phi=0$ at $a^{-1}\simeq 1{\rm GeV}$. 
We find ``constituent scalar-quark/quark picture" 
for both scalar-quark hadrons and chimera hadrons.
Namely, the mass of the new-type hadron composed of 
$m$ $\phi$'s and $n$ $\psi$'s, $M_{{m}\phi+{n}\psi}$, 
approximately satisfies $M_{{m}\phi+{n}\psi}\simeq  {m} M_\phi +{n} M_\psi$, 
where $M_\phi$ and $M_\psi$ are the constituent scalar-quark and quark masses, 
respectively. 
We estimate the constituent scalar-quark mass $M_\phi$ 
for $m_\phi=0$ at $a^{-1}\simeq 1{\rm GeV}$ 
as $M_\phi \simeq$ 1.5-1.6GeV, 
which is much larger than the constituent quark mass 
$M_\psi\simeq 400{\rm MeV}$ in chiral limit. 
Thus, 
scalar-quarks acquire a large mass due to large quantum corrections 
by gluons in the systems including scalar-quarks.
Together with other evidences of mass generation of 
glueballs and charmonia, 
we conjecture that all colored particles generally acquire 
a large effective mass due to dressed gluon effects. 
In addition, 
the large mass generation of point-like colored scalar particles 
indicates that plausible diquarks used in effective hadron models 
cannot be described as the point-like 
particles and should have a much larger size than 
$a \simeq 0.2{\rm fm}$. 
\end{abstract}

\maketitle

\section{Introduction}
\label{sec1}
The origin of mass is one of the fundamental and fascinating subjects 
in physics for a long time. 
A standard interpretation of mass origin is the Yukawa interaction 
with the Higgs field \cite{H64}, 
of which observation is one of the most important aims of 
the Large Hadron Collider (LHC) experiment at CERN. 
However, the mass of Higgs origin is only about 1\% 
of total mass in the world,
because the Higgs interaction only provides  
the current quark mass (less than 10MeV for light quarks) 
and the lepton mass (0.51MeV for electrons) \cite{PDG}.
On the other hand, apart from unknown dark matter, 
about 99\% of mass of matter in the world originates 
from the strong interaction, 
which actually provides the large constituent quark mass 
$M_\psi =(300-400){\rm MeV}$ \cite{RGG75}. 
Such a dynamical fermion-mass generation in the strong interaction 
can be interpreted as spontaneous breaking of chiral symmetry, 
which was first pointed out by Y.~Nambu~et~al. \cite{NJ61}. 
According to the chiral symmetry breaking, light quarks are considered to have 
a large constituent quark mass of about $400{\rm MeV}$ \cite{H84,M83}. 
In terms of the mass-generation mechanism, 
it is interesting to investigate high density and/or high temperature 
systems, because 
chiral symmetry is expected to be restored in 
the system \cite{HK94,KSWGSSS83,IOS05}. 
Several theoretical studies suggest that such a restoration 
 affects the hadron nature \cite{HK94,DK87,HL92,IOS05}. 
Actually, the experiments at  CERN SPS \cite{CERES} and KEK \cite{KEK}
indicate the change of nature of the vector mesons $\rho$, $\omega$ and $\phi$ 
in finite density system. 
These experiments are important for the study of the mass generation 
originated from chiral symmetry breaking. 
In this way, the origin of mass and its related topics, 
including the Higgs-particle search at LHC 
and chiral symmetry breaking in QCD, 
are energetically studied in both sides of theories and experiments.

Then, a question arises: Is there any mechanism of dynamical mass 
generation without chiral symmetry breaking? 
To answer this question, we note the following examples 
supporting the existence of such a mechanism. 
One example is gluons, i.e., colored vector particles in QCD.
While the gluon is massless in perturbation QCD, 
non-perturbative effects due to the self-interaction of gluons 
seem to generate a large effective gluon mass as $(0.5-1.0){\rm GeV}$, 
which is measured in lattice QCD \cite{MO87,AS99}. 
Actually, glueballs, which are composed of gluons, 
have a large mass, e.g., about $1.5{\rm GeV}$ \cite{MP99ISM02, ISM02} 
even for the lightest glueball ($J^{PC}=0^{++}$), 
and the scalar meson $f_0(1500)$ is one of the candidate 
of the lowest $0^{++}$ glueball in experimental side \cite{PDG}. 
Another example is charm quarks. 
The current mass of charm quarks is about $1.2{\rm GeV}$ 
at the renormalization point $\mu=1{\rm GeV}$ \cite{PDG}. 
In the quark model, however, the constituent charm-quark mass 
which reproduces masses of charmonia is set 
to be about $1.6{\rm GeV}$ \cite{RGG75}. 
The about 400MeV difference between the current and the constituent 
charm-quark masses can be explained as dynamical mass generation 
without chiral symmetry breaking, since there is no chiral symmetry 
for such a heavy-quark system. 

These examples suggest that there is another type of mass generation 
without chiral symmetry breaking and the Higgs mechanism. 
Then, we conjecture that, even without chiral symmetry breaking, 
large dynamical mass generation 
generally occurs in the strong interaction, i.e., 
{\it all colored particles 
have a large effective mass generated by dressed gluon effects.} 
This is one of the main motivations of the present study 
on the mass generation of the hadronic systems 
including light scalar-quarks, which do not have the chiral symmetry.

The study is also motivated by the ``diquark picture'' of hadrons 
\cite{CR81, FJL82, P76, W06}. 
The diquark is made of two quarks which are strongly correlated in 
color anti-triplet $\bar {\bf 3}$ channel, 
where the one-gluon-exchange interaction between two quarks is most attractive.
Sometimes, diquarks are treated as point-like objects or 
local boson fields. 
Degrees of freedom of diquarks seems to be 
important in some aspects of hadron physics. 
For example, nucleon structure functions can be explained by the model 
in which nucleon is the 
admixture of a diquark and a quark \cite{CR81,FJL82}. 
Diquarks are also important for hadron spectra \cite{P76,W06}. 
In an extremely high density system, 
diquark condensation is considered to appear due to the 
attractive interaction of one-gluon exchange in color $\bar {\bf 3}$ channel. 
Such a phase is called color superconductivity phase \cite{BL84,ARW99} and 
it may exist in the interior of neutron stars. 
However, the properties of diquarks such as 
the size and the mass are not well understood,
in spite of the frequent use of diquarks in hadron physics. 
We note that point-like limit of diquarks in color $\bar {\bf 3}$ channel 
can be regarded as the scalar-quark in fundamental representation of SU(3)$_c$ 
in scalar QCD \cite{CL}. 
Therefore, the theoretical study of scalar-quarks with a reliable method is 
expected to be a good test place to check the validity of the point-like 
diquark picture.

In this paper, according to these motivations, 
we study the light scalar-quark belonging to 
the fundamental representation of SU(3)$_c$, 
and their color-singlet hadronic states 
using quenched SU(3)$_c$ lattice QCD \cite{IST06}. 
We investigate ``scalar-quark hadrons'', which are the color-singlet 
bound states of scalar-quarks $\phi$, 
and ``chimera hadrons'', which are the bound states of $\phi$ 
and (ordinary) quarks $\psi$. 
Using the standard technique of the lattice gauge theory, 
we calculate the temporal correlators of these new-type hadrons 
at the quenched level, and extract their mass of 
the lowest energy state from the correlators. 
As a result, we find a large mass of the hadron including 
scalar-quarks $\phi$. Also we find the ``constituent scalar-quark/quark 
picture'' for all the new-type hadrons. Namely, the mass of 
the new-type hadron which is composed of $m$ $\phi$'s and $n$ $\psi$'s, 
$M_{m\phi+n\psi}$, satisfies $M_{m\phi+n\psi}\simeq m M_\phi +n M_\psi$, 
where $M_\phi$ and $M_\psi$ are the constituent scalar-quark and quark masses, 
respectively. 
The constituent scalar-quark mass $M_\phi$ estimated from the masses 
of these new-type hadrons is found to be much larger than 
the constituent quark mass $M_\psi \simeq 400{\rm MeV}$ for light quarks.
This result suggests that, in the system of colored scalar particles, 
there occurs large mass generation in the strong interaction 
without chiral symmetry breaking. 

The article is organized as follows. 
In Sec.~\ref{sec2}, the new states including scalar-quarks, 
``scalar-quark hadrons'' and ``chimera hadrons'', are introduced. 
In Sec.~\ref{sec3}, we present the formalism of lattice QCD 
including scalar-quarks, the setup of the lattice calculations, 
and the method to extract the masses of the new-type hadrons. 
In Sec.~\ref{sec4}, we show the numerical results of 
the scalar-quark hadrons, i.e., 
scalar-quark mesons $\phi^\dagger \phi$ and 
scalar-quark baryons $\phi\phi\phi$, and investigate 
the constituent scalar-quark picture and mass generation of scalar-quarks.  
In Sec.~\ref{sec5}, we show the numerical results of the chimera hadrons, 
i.e., chimera mesons $\phi^\dagger \psi$ and chimera baryons $\psi\psi\phi$ 
and $\phi\phi\psi$, and analyse them in terms of the 
constituent scalar-quark/quark picture and their structure.
Sec.~\ref{sec6} is devoted to conclusion and discussion.

\begin{table*}[ht]
\caption{Summary table of new-type hadrons, i.e., scalar-quark hadrons and 
chimera hadrons. Names, Lorentz properties and gauge-invariant local operators 
used for our lattice calculation are listed. 
The indices $i,j$ and $k$ denote the scalar-quark 
flavor degrees of freedom. $\Gamma^{ij}_M$, $\Gamma^{ijk}_B$ and 
$\Gamma^{ij}_B$ are some tensors on the scalar-quark flavor.}
\label{tab1}
\begin{tabular}{cccl}
\hline\hline
Names & \ & Lorentz properties &  \ \ \ \ \ \ \ \ Local operators \\
\hline
Scalar-quark meson & ($\phi^\dagger\phi$) & Scalar 
& $M_s(x)= \Gamma^{ij}_M \phi^{\dagger i}_a (x)\phi_a^j(x)$\\
Scalar-quark baryon & ($\phi\phi\phi$)  & Scalar 
& $B_s(x)= \Gamma^{ijk}_B \epsilon_{abc}\phi_a^i (x)\phi_b^j (x)\phi_c^k (x)$\\
Chimera meson & ($\phi^\dagger\psi$) & Spinor 
& $C_M^\alpha(x)= \phi^\dagger_a (x) \psi_a^\alpha (x)$\\
Chimera baryon & ($\psi\psi\phi$) & Scalar 
& $C_{B}(x)= \epsilon_{abc} ({\psi_a}^T(x)C\gamma_5\psi_b(x)) \phi_c (x)$\\
Chimera baryon & ($\phi\phi\psi$) & Spinor & $C_{B}^{\alpha}(x)= 
\Gamma^{ij}_B\epsilon_{abc}\phi_a ^i(x)\phi_b^j (x)\psi_c^\alpha (x)$\\
\hline\hline
\end{tabular}
\vspace{-0.4cm}
\end{table*}

\section{New states including scalar-quarks: scalar-quark hadrons 
and chimera hadrons }
\label{sec2}

In this section, we explain the new-type hadrons introduced in this study: 
various color-singlet systems including light scalar-quarks $\phi$.
We call the color-singlet state composed only of scalar-quarks $\phi$ as 
a ``scalar-quark hadron'': 
the scalar-quark meson is composed of a scalar-quark and an anti-scalar-quark, 
 $\phi^\dagger \phi$, and the scalar-quark baryon 
is made of three scalar-quarks, $\phi\phi\phi$. 
The scalar-quark meson field $M_s(x)$ and 
the scalar-quark baryon field $B_s(x)$ can be expressed by gauge-invariant 
local operators as 
\begin{eqnarray}
&&M_s(x)= \Gamma^{ij}_M\phi_a^{\dagger i} (x) \phi_a^j(x)
\label{SQM},\\
&&B_s(x)= \Gamma^{ijk}_B \epsilon_{abc}\phi_a^i(x)\phi_b^j(x)\phi_c^k(x),
\label{SQB}
\end{eqnarray}
where the scalar-quark field $\phi_a(x)$ ($a$=1,2,3) 
belongs to the fundamental (${\bf 3}_c$) representation of SU(3)$_c$.
In above equations, $a,b$ and $c$ denote the color indices, 
and $\epsilon_{abc}$ is the antisymmetric tensor for color indices.
Here, we have generally introduced different species of scalar-quarks, 
which leads to the ``scalar-quark flavor" degrees of freedom. 
In Eqs.(\ref{SQM}) and (\ref{SQB}), 
$i,j$ and $k$ denote the scalar-quark flavor indices, 
and $\Gamma^{ij}_M$ and $\Gamma^{ijk}_B$ are some tensors 
on the scalar-quark flavor. 
We note that, for the local-operator description of scalar-quark baryons, 
the number $N_f^{\rm scalar}$ of the scalar-quark flavor  
should be $N_f^{\rm scalar} \ge 3$. 
Actually, for $N_f^{\rm scalar} \le 2$, the local scalar-quark baryon operator 
vanishes due to the antisymmetric tensor $\epsilon_{abc}$, 
although a non-local operator description is possible 
for scalar-quark baryons. 
In contrast, scalar-quark mesons can be described with local operators 
at any number of $N_f^{\rm scalar} (\ge 1)$. 

We call the color-singlet state made of scalar-quarks $\phi$ and 
(ordinary) quarks $\psi$ as a ``chimera hadron'': 
chimera mesons $\phi^\dagger \psi$ and chimera baryons, 
$\psi\psi\phi$ and  $\phi\phi\psi$. 
The chimera meson field $C_M^\alpha (x)$ and the chimera baryon field, 
$C_B(x)$ and $C_B^\alpha(x)$, can be expressed by gauge-invariant 
local operators as 
\begin{eqnarray}
&&C_M^\alpha (x)= \phi_a^\dagger(x)\psi_a^\alpha(x),\\
&&C_B(x)=\epsilon_{abc}(\psi^T_a(x)C\gamma_5\psi_b(x))\phi_c(x),\\
&&C_B^\alpha(x)
=\Gamma^{ij}_B\epsilon_{abc}\phi_a^i(x)\phi_b^j(x)\psi_c^\alpha(x),
\label{CB}
\end{eqnarray}
where $\alpha$ denotes the spinor index. 
$\Gamma^{ij}_B$ is some tensor on the scalar-quark flavor. 
We also note that, for the local-operator description of 
the $\phi\phi\psi$-type chimera baryon, 
the scalar-quark number $N_f^{\rm scalar}$ is to be $N_f^{\rm scalar} \ge 2$. 
For $N_f^{\rm scalar} = 1$, the local operator of the $\phi\phi\psi$-type 
chimera baryon vanishes due to the antisymmetric tensor $\epsilon_{abc}$ 
like the scalar-quark baryon case, 
although a non-local operator description is possible. 
On the other hand, 
chimera mesons $\phi^\dagger\psi$ and chimera baryons $\psi\psi\phi$ 
can be described with local operators at any number of 
$N_f^{\rm scalar} (\ge 1)$. 

We comment on the single-flavor case of the scalar-quark. 
In this case, from the symmetric property of bosons $\phi$ and 
the anti-symmetric property on the color quantum number, 
the space-time coordinates of bosons $\phi$ are to be 
anti-symmetrized in scalar-quark baryons $\phi\phi\phi$ 
and chimera baryons $\phi\phi\psi$, and therefore 
the local operators for these hadrons inevitably vanish, 
as explained above.
In fact, scalar-quark baryons and chimera baryons 
are expressed with anti-symmetric functions $\chi(x_i)$ as 
\begin{eqnarray}
B_s&\sim&\epsilon_{abc}\sum_{x_1,x_2,x_3}
\chi(x_1, x_2, x_3)\phi_a(x_1)\phi_b(x_2)\phi_c(x_3),~~~ \\
C_B^\alpha&\sim&\epsilon_{abc}\sum_{x_1,x_2}
\chi(x_1, x_2)\phi_a(x_1)\phi_b(x_2)\psi^\alpha_c,
\end{eqnarray}
which are not local form and should be constructed 
in a gauge-invariant manner \cite{TS0102,OST05}.
However, the states anti-symmetrized on the space-time coordinates of $\phi$
are expected to be rather heavy, 
since the one or two scalar-quarks in those hadrons 
are not in the lowest S-wave state in a single-particle picture. 
(This situation is similar to the early stage of the quark model 
and the introduction of ``color'' quantum number for quarks 
by Han and Nambu \cite{HN65}.)
Therefore, we consider the multi-flavor case 
and use the local operators for 
$\phi\phi\phi$ and $\phi\phi\psi$ in Eqs.(\ref{SQB}) and (\ref{CB}), 
since these operators are expected to couple with 
the lowest state in each channel, 
which is suitable for the investigation of 
the mass generation of scalar-quarks. 

\begin{figure}[ht] 
\includegraphics[width=8cm]{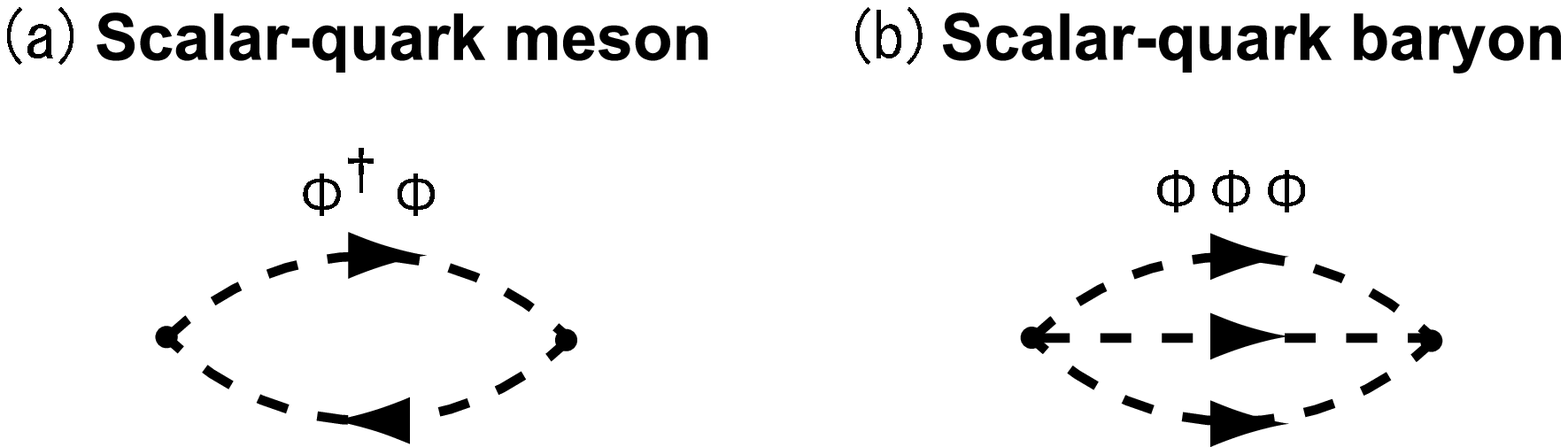} 
\caption{Diagrams for scalar-quark hadrons, corresponding to 
(a) the two-point correlator of scalar-quark mesons ($\phi^\dagger \phi$) 
and (b) that of scalar-quark baryons ($\phi\phi\phi$). 
The dotted line denotes the scalar-quark $\phi$, and the arrow 
denotes the current of the color in fundamental representation.
As for the scalar-quark meson, 
we only consider the connected diagram, i.e., 
non-singlet states of the scalar-quark flavor. 
For scalar-quark baryons, no disconnected diagram appears 
because of the color conservation.
}
\label{fig1}
\end{figure}

\begin{figure}[ht]
\includegraphics[width=8cm]{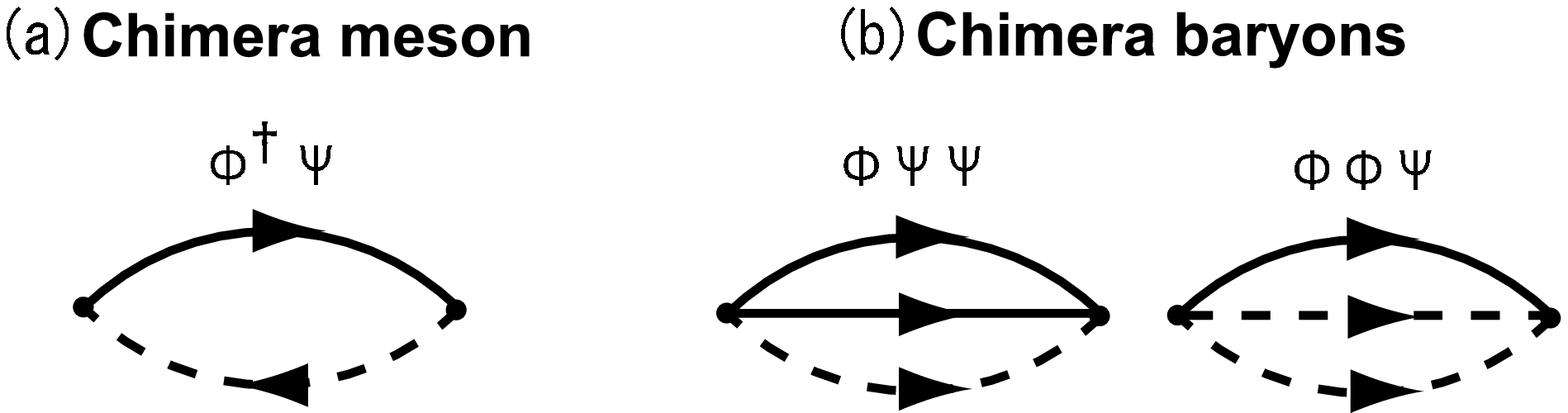}
\caption{Diagrams for chimera hadrons corresponding to 
(a) the two-point correlator of chimera mesons ($\phi^\dagger \psi$) 
and (b) that of chimera baryons ($\psi\psi\phi$ and $\phi\phi\psi$). 
The dotted line corresponds to the scalar-quark $\phi$ and the solid line 
corresponds to the (ordinary) quark $\psi$. 
For chimera mesons and chimera baryons, 
no disconnected diagram appears 
because of the color and the quark-number conservation.
}
\label{fig2}
\end{figure}

In Fig.~1, we show the diagrams for scalar-quark hadrons, 
corresponding to the two-point correlator of scalar-quark mesons 
($\phi^\dagger \phi$) and that of scalar-quark baryons ($\phi\phi\phi$).
The dotted line denotes the scalar-quark $\phi$, and the arrow 
denotes the current of the color in fundamental representation. 
For scalar-quark mesons, we only investigate 
non-singlet states of the scalar-quark flavor, and 
therefore the disconnected diagram 
does not appear in our calculation. 
For scalar-quark baryons, 
no disconnected diagram appears at the quenched level 
like ordinary baryons,
because of the conservation law of the color quantum number. 
(Note that the scalar-quark $\phi$ belongs to 
the fundamental representation, ${\bf 3}_c$.)

In Fig.~2, we show the diagrams for chimera hadrons, corresponding to 
the two-point correlator of the chimera meson ($\phi^\dagger \psi$) and 
that of the chimera baryon ($\psi\psi\phi$ and $\phi\phi\psi$). 
The dotted line corresponds to the scalar-quark $\phi$ and 
the solid line corresponds to the (ordinary) quark $\psi$. 
Also for chimera mesons and chimera baryons, 
no disconnected diagram appears at the quenched level,
because of the conservation law of the color quantum number 
and the quark number.
In fact, disconnected diagrams appears only 
for the flavor-singlet scalar-quark meson.

We note that the statistical property for these new-type 
hadrons is rather different from that for ordinary hadrons. 
For example, the scalar-quark baryon $\phi\phi\phi$ is 
a boson and the chimera meson $\phi^\dagger\psi$ is a fermion, 
while ordinary baryons are fermions and ordinary mesons are bosons. 
In Table \ref{tab1}, we summarize the names, Lorentz properties 
and gauge-invariant local operators for the new-type hadrons. 

We study the mass generation of scalar-quarks 
by investigating the various color-singlet systems 
including scalar-quarks in the lattice gauge theory.

\section{Formalism and setup of lattice QCD for scalar-quarks}
\label{sec3}

In this section, we present the lattice QCD formalism for scalar-quarks 
and the setup for the actual calculation. 
To include scalar-quarks $\phi$ together with quarks $\psi$ and 
gluons $A_\mu\equiv A_\mu^a T^a$ in QCD,  
we adopt the generalized QCD Lagrangian density 
\begin{eqnarray}
\cal{L}&=& -\frac{1}{4}G^a_{\mu\nu}G^{a\mu\nu} + \cal{L}_{\rm F} 
+\cal {L}_{\rm SQ}, \\
{\cal L}_{\rm F} &=& \bar\psi (i \Slash{D} - m_\psi) \psi, \\
\cal{L}_{\rm SQ}&=&{\rm tr} \ (D_\mu \phi)^\dagger(D^\mu\phi)-m_{\phi}^2 
\ {\rm tr} \ \phi^\dagger\phi,
\end{eqnarray}
in the continuum limit in the Minkowski space.
Here, the scalar-quark field $^t\phi=(\phi_1,\phi_2,\phi_3)_c$ belongs to 
the fundamental ({\bf 3}$_c$) representation of the color SU(3) 
like the quark field $^t\psi=(\psi_1,\psi_2,\psi_3)_c$.
$m_\psi$ is the bare mass of quarks and is expressed 
as a diagonal matrix in the flavor space. 
Similarly, $m_\phi$ is the bare mass of scalar-quarks, 
and is expressed as a diagonal matrix in the scalar-quark flavor space. 
In this study, 
all the quarks are taken to have the same bare mass $m_\psi$, and 
all the scalar-quarks are taken to have 
the same bare scalar-quark mass $m_\phi$. 

$G^a_{\mu\nu}$ denotes the field strength tensor defined as 
\begin{eqnarray}
G^a_{\mu\nu}\equiv \partial_\mu A^a_\nu-\partial_\nu A^a_\mu 
+gf^{abc} A_\mu^b A_\nu^c,
\end{eqnarray}
where $f^{abc}$ is the structure constant of SU(3)$_c$ and 
$\Slash D\equiv \gamma^\mu D_\mu$ denotes the covariant derivative,
\begin{eqnarray}
D_\mu&\equiv& \partial_\mu -ig A_\mu. 
\end{eqnarray}

In the lattice calculation, we discretize this action in the Euclidean metric. 
For the gluon sector, we adopt the standard plaquette action \cite{Rothe}, 
\begin{eqnarray}
S_{\rm G}\equiv \frac{\beta}{N_c}\sum_{x,\mu,\nu}
 {\rm Re}{\rm Tr}\{1-P_{\mu\nu}(x)\},
\end{eqnarray}
with $\beta\equiv 2N_c/g^2$. The plaquette $P_{\mu\nu}(x)$ is given by 
\begin{eqnarray}
P_{\mu\nu}(x) \equiv U_\mu(x) U_\nu(x+\hat\mu) 
U_\mu^\dagger (x+\hat \nu) U_\nu^\dagger (x),
\end{eqnarray}
where $U_\mu(x) \in {\rm SU}(N_c)$ is the link variable, 
corresponding to the gluon field as 
$U_\mu(x) \simeq \exp(iagA_\mu(x))$ near the continuum limit. 
$\hat \mu$ denotes the unit vector of the direction 
$\mu$ on the lattice.

For the quark part, we use the Wilson fermion action \cite{Rothe},
\begin{eqnarray}
S_F&\equiv& \sum_{x,y} \bar\psi(x)K(x,y)\psi(y), \nonumber \\
K(x,y)&\equiv& \delta_{x,y}-\kappa\sum_\mu 
\{ ({\bf 1}-\gamma_\mu)U_\mu(x)\delta_{x+\hat\mu,y} \nonumber\\
&&+({\bf 1}+\gamma_\mu)U_\mu^\dagger(y)\delta_{x,y+\hat\mu} \},
\end{eqnarray}
where $\kappa$ is the hopping parameter related to 
the bare quark mass $m_\psi$. 
The Wilson fermion action includes $O(a)$ discretization error,
which can be reduced by using the $O(a)$-improved fermion action 
\cite{EKM97,Iida0605}. 

For scalar-quarks, we take the local and simplest gauge-invariant action, 
\begin{eqnarray}
S_{\rm SQ}\equiv \sum_{x,y}{\phi}^\dagger(x)
\{-\sum_\mu(\delta_{{x+\hat\mu},y}U_\mu (x)+\delta_{{x-\hat\mu},y}
{U_\mu}^\dagger(y)\nonumber \\
 -2\delta_{xy}{\bf 1})+m_{\phi}^2\delta_{xy}{\bf 1}\} {\phi}(y).
\end{eqnarray} 
Note that the scalar-quark action only includes 
$O(a^2)$ discretization error, 
which is the same order of that in the gauge action and is 
higher order than that of the Wilson fermion action. 
In contrast to the fermion case, the indirect bare-mass parameter 
such as $\kappa$ for quarks is not needed to calculate the scalar-quark system,
since no doubler appears and the Wilson term is not necessary for 
the scalar field. Therefore, we can directly use the bare mass $m_\phi$ 
for the lattice calculation of scalar-quarks. 

In this lattice action, the scalar-quark $\phi$ is introduced as a 
local field on the lattice with spacing $a$, 
and therefore the scalar-quark $\phi$ can be regarded as 
an elementary local field at the renormalization point $a^{-1}$ or 
a point-like (composite) object with the intrinsic size of $O(a)$.
When the scalar-quark $\phi$ is identified as a diquark, 
the scalar-quark $\phi$ should be interpreted 
as an effective degrees of freedom 
like an idealized point-like diquark with the size of $O(a)$.

Here, we comment on the bare mass of the scalar-quark.
As is well-known, the current quark mass is induced by the interaction 
with the Higgs scalar field, but appears as a bare parameter in QCD. 
Similarly, the bare mass $m_\phi$ of the scalar-quark 
can be regarded as an effective mass 
at the scale of $a$ in the lattice gauge theory, 
and $m_\phi$ may be induced by an underlying interaction except for QCD,
e.g., $m_\phi$ may appear as the mean-field value in the $\phi^4$ theory.
In fact, 
once the scalar-quark field $\phi$ with the bare mass $m_\phi$ 
at the renormalization scale $a$ is given, 
the effect of the gluon interaction during the propagation of 
scalar-quarks in the gluonic medium  
can be investigated in the quenched lattice gauge theory. 

We choose $\beta\equiv 2N_c/g^2=5.70$, which corresponds to 
the lattice spacing $ a \simeq 0.18{\rm fm} \simeq (1.1{\rm GeV})^{-1}$ 
\cite{TS0102}.
One of the reason to adopt the coarse lattice $a \simeq 0.2{\rm fm}$ is 
to investigate the idealized point-like diquark 
with a possible extension of about 0.2fm.
We adopt $16^3\times 32$ lattice, which corresponds to the spatial volume 
$V\simeq(2.9{\rm fm})^3$. 
The number of the gauge configurations $N_{\rm conf}$ is 100. 
We take the configurations every 500 sweeps after 20,000 sweeps 
for thermalization.
For the bare scalar-quark mass, 
$m_\phi= 0.0, 0.11, 0.22$ and $0.33{\rm GeV}$ are adopted. 
For the bare quark mass, we take the hopping parameters 
$\kappa=0.1650, 0.1625$ and $0.1600$. 
The corresponding bare quark masses 
$m_{\psi}$ are roughly estimated as 
$0.09, 0.14$ and $0.19{\rm GeV}$ \cite{BCSVW94, TUOK05}, 
if one uses the tree-level relation $2m_\psi a=1/\kappa-1/\kappa_c$, 
where $\kappa_c = 0.1694$ is the critical hopping parameter 
where pion is massless. 
These parameters are summarized in Table \ref{tab2}. 
In Table \ref{tab3}, 
the masses of pions, $\rho$ mesons and nucleons 
calculated on our lattice are shown 
for each hopping parameter $\kappa$, together with 
the rough estimate of bare quark masses $m_\psi$. 
In this paper, we adopt the jackknife error estimate 
for the lattice results.

\begin{table*}[ht]
\caption{Parameters and relevant quantities 
in the lattice QCD calculation. 
The lattice spacing $a^{-1}$ is set by the string tension 
$\sigma\simeq (427{\rm MeV})^2$ \cite{TS0102}. }
\label{tab2}
\begin{tabular}{cccccc}
\hline\hline
$\beta$ &Lattice size& $a^{-1}$& $N_{\rm conf}$ &
bare scalar-quark mass $m_{\phi}$ & $\kappa$\\ 
\hline
$5.70$ & $16^3\times 32$&  1.1 GeV & 100 
& 0.00, 0.11, 0.22, 0.33GeV
& 0.1650, 0.1625, 0.1600 \\
\hline\hline
\end{tabular}
\end{table*}

\begin{table*}[ht]
\caption{Masses of $\pi$ (pseudoscalar meson), $\rho$ (vector meson) and 
N (nucleon) calculated on the lattice for each hopping parameter $\kappa$. 
The error estimate is done by the jackknife method. 
$\chi^2$ per degrees of freedom $N_{\rm df}$ and 
the rough estimate of the bare quark mass $m_\psi$ are added.
The unit is GeV.
}
\label{tab3}
\begin{tabular}{ccccc}
\hline\hline
$\kappa$ & $M_{\pi}$ ($\chi^2/N_{\rm df}$) &  
$M_{\rho}$ ($\chi^2/N_{\rm df}$)&  
$M_{\rm N}$ ($\chi^2/N_{\rm df}$) & $m_\psi$ \\
\hline
0.1650 & 0.4926$\pm$0.0036 (0.748) & 0.7177$\pm$0.0091 (2.772) 
& 1.1551$\pm$0.0021 (1.426) & 0.087\\
0.1625 & 0.6303$\pm$0.0035 (0.599) & 0.7969$\pm$0.0063 (1.462) 
& 1.2934$\pm$0.0017 (0.627) & 0.14\\
0.1600 & 0.7510$\pm$0.0036 (0.715) & 0.8791$\pm$0.0053 (0.955) 
& 1.4453$\pm$0.0013 (0.817) &0.19\\
\hline\hline
\end{tabular}
\end{table*}

\begin{table*}[ht]
\caption{
The mass of scalar-quark hadrons: scalar-quark 
mesons $\phi^\dagger\phi$ and scalar-quark 
baryons $\phi\phi\phi$ in term of the bare scalar-quark mass 
$m_{\phi}$ at $a^{-1} \simeq 1.1{\rm GeV}$.
$M_{\phi^\dagger\phi}$ and $M_{\phi\phi\phi}$ 
denote their mass. The unit of mass is GeV.
$\chi^2$ over degrees of freedom $N_{\rm df}$ and 
fit ranges are 
also listed. The error estimate is done by the jackknife method.}
\begin{tabular}{ccc}
\hline\hline
$m_{\phi}$ & $ \ M_{\phi^\dagger \phi}$ ($\chi^2/N_{\rm df}$, [fit range]) 
 & \ $M_{\phi\phi\phi}$ ($\chi^2/N_{\rm df}$, [fit range]) \\
\hline
0     & \ 3.016$\pm$0.0039 (0.875, [7-14]) 
& \ 4.697$\pm$0.0096 (0.403, [7-15])\\
0.11 & \ 3.025$\pm$0.0039 (0.878, [7-14]) 
& \ 4.711$\pm$0.0096 (0.407, [7-15])\\
0.22 & \ 3.052$\pm$0.0039 (0.886, [7-14]) 
& \ 4.751$\pm$0.0095 (0.417, [7-15])\\
0.33 & \ 3.095$\pm$0.0038 (0.899, [7-14]) 
& \ 4.818$\pm$0.0094 (0.435, [7-15])\\
\hline\hline
\end{tabular}
\label{tab4}
\end{table*}

We calculate the following temporal correlator $G(t)$ of 
the hadronic operator $O(x,t)$,
\begin{eqnarray}
G(t)\equiv \frac{1}{V}\sum_{\vec x} \langle O(\vec x,t) O^{\dagger}(\vec 0,0)
\rangle,
\end{eqnarray} 
where the total momentum is projected to be zero. 
To extract the lowest-state mass $M_0$ of a hadron, 
we fit the temporal correlator by the single exponential function 
$G_{\rm fit}(t)\equiv A\exp (-M_0 t)$ 
 in the region $[t_{\rm min}, t_{\rm max}]$ where the lowest energy state dominates. 
For the determination of the fit range, 
we construct so-called effective mass, 
\begin{eqnarray}
m_{\rm eff}(t)\equiv \ln\frac{G(t)}{G(t+1)}.
\end{eqnarray}
If $G(t)$ is dominated by the lowest state with mass $M_0$ 
in a certain region for large $t$, 
then $m_{\rm eff}(t)\simeq M_0$ for the $t$ region. 
In other words, if there is a plateau region of $m_{\rm eff}(t)$ for 
large $t$, the correlator is dominated by the lowest energy state for 
the $t$ region. 
Therefore, we set the fit region $[t_{\rm min}, t_{\rm max}]$ 
of the correlator from the plateau region 
$[t_{\rm min}, t_{\rm max}-1]$ of the effective mass. 
\begin{figure}[h]
\begin{center}
\begin{tabular}{cc}
\includegraphics[width=4.4cm]{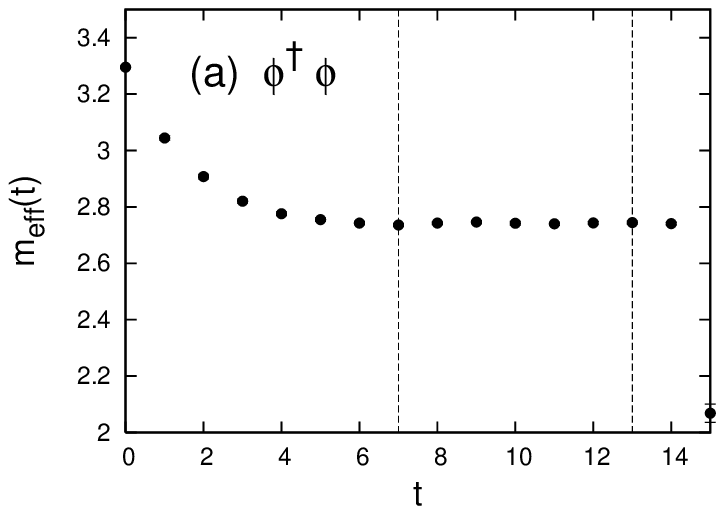} 
\hspace{-0.4cm} 
&
\includegraphics[width=4.4cm]{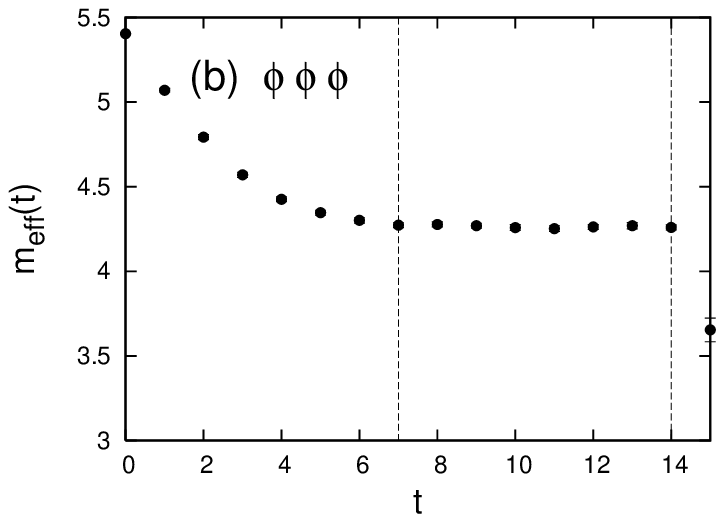} \\
\includegraphics[width=4.4cm]{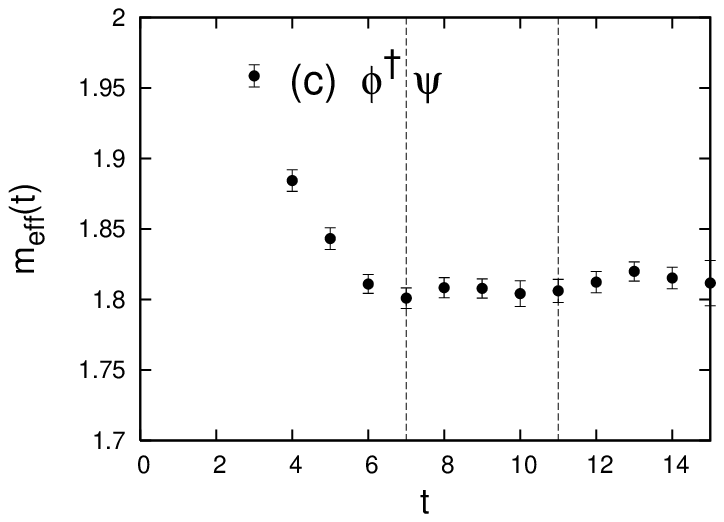} 
\hspace{-0.4cm} 
&
\includegraphics[width=4.4cm]{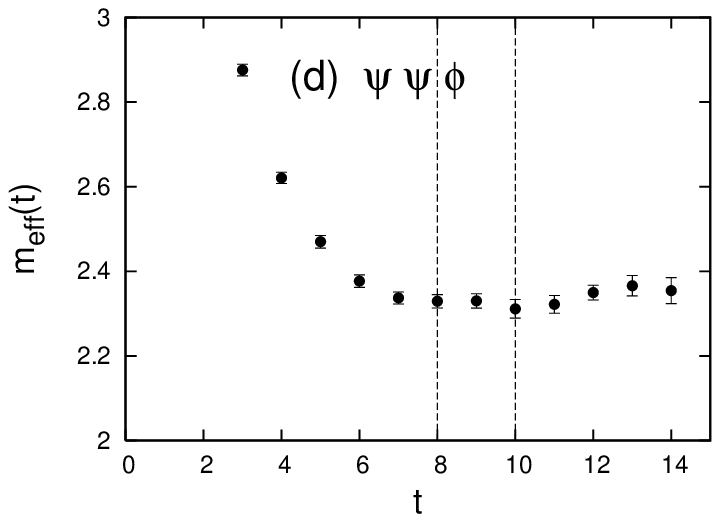} \\
\multicolumn{2}{c}{\includegraphics[width=4.4cm]{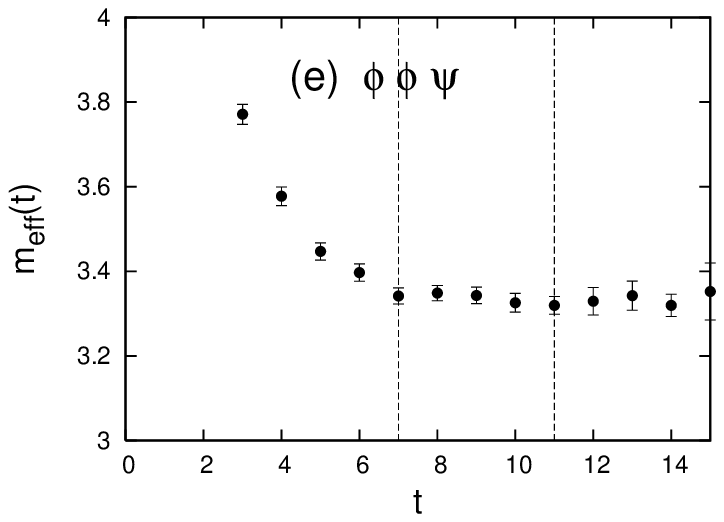}}
\end{tabular}
\caption{Typical examples of 
effective mass plots for (a) scalar-quark mesons 
$\phi^\dagger\phi$, (b) scalar-quark baryons $\phi\phi\phi$, 
(c) chimera mesons $\phi^\dagger\psi$, 
(d) chimera baryons $\psi\psi\phi$ and (e) chimera baryons $\phi\phi\psi$. 
In the figure, the bare scalar-quark mass $m_\phi$ 
for scalar-quark hadrons and chimera hadrons 
is $m_\phi=0$ and the pion mass $M_\pi$ for chimera hadrons is 
$M_\pi=0.75{\rm GeV}$. 
The unit of horizontal axis is lattice spacing $a$ and that of 
vertical axis is $a^{-1}$.
For large $t$, there is a plateau region and the correlator $G(t)$ 
for the scalar-quark meson 
is fitted in the corresponding region by single exponential function. 
The vertical dashed lines denote the fit ranges of corresponding correlators. 
%
}
\label{fig3}
\end{center}
\end{figure}
%
%
Figure 3 is the examples of effective masses for (a)
scalar-quark mesons $\phi^\dagger \phi$, 
(b) scalar-quark baryons $\phi\phi\phi$, 
(c) chimera mesons $\phi^\dagger \psi$, 
(d) chimera baryons $\psi\psi\phi$ and 
(e) chimera baryons $\phi\phi\psi$. 
In these effective masses, we can see the plateau regions. 
We note here that error estimate is done by the jackknife method for 
all the data in this study. 

In this way, using the standard lattice QCD technique, 
we can extract the lowest mass for 
the new-type hadrons including scalar-quarks 
like the ordinary hadrons.

\section{Results for scalar-quark hadrons}
\label{sec4}

In this section, we show lattice results for the mass of scalar-quark hadrons, 
namely, scalar-quark mesons $\phi^\dagger \phi$ and scalar-quark baryons 
$\phi\phi\phi$. 
Figure 4(a) shows the scalar-quark meson mass $M_{\phi^\dagger \phi}$ 
plotted against the bare scalar-quark mass $m_{\phi}$. 
In Fig.~4(a), we find a large mass of about 3GeV 
for the scalar-quark meson, even for zero bare scalar-quark mass, $m_\phi=0$. 
This scalar-quark meson mass is rather large 
compared to the ordinary low-lying mesons composed of light quarks. 
Figure 4(b) shows the scalar-quark baryon mass $M_{\phi\phi\phi}$ plotted 
against $m_{\phi}$. 
In Fig.~4(b), we find also a large mass of about 4.7GeV 
for the scalar-quark baryon, even for zero bare scalar-quark mass. 
The numerical data for these figures are listed in Table \ref{tab4}. 

From the data of these scalar-quark hadrons, 
the ``constituent scalar-quark picture'' seems to be satisfied, 
i.e., $M_{\phi^\dagger \phi}\simeq 2M_{\phi}$ 
and $M_{\phi\phi\phi}\simeq 3M_{\phi}$, where $M_{\phi}$ is the constituent 
(dynamically generated) scalar-quark mass and is 1.5-1.6GeV. 
Thus, in the scalar-quark hadron systems, 
there occurs large mass generation of scalar-quarks 
(large quantum correction to the bare scalar-quark mass) 
in the strong interaction without chiral symmetry breaking. 

\begin{figure}[ht]
\begin{center}
\begin{tabular}{cc}
\includegraphics[width=4.5cm]{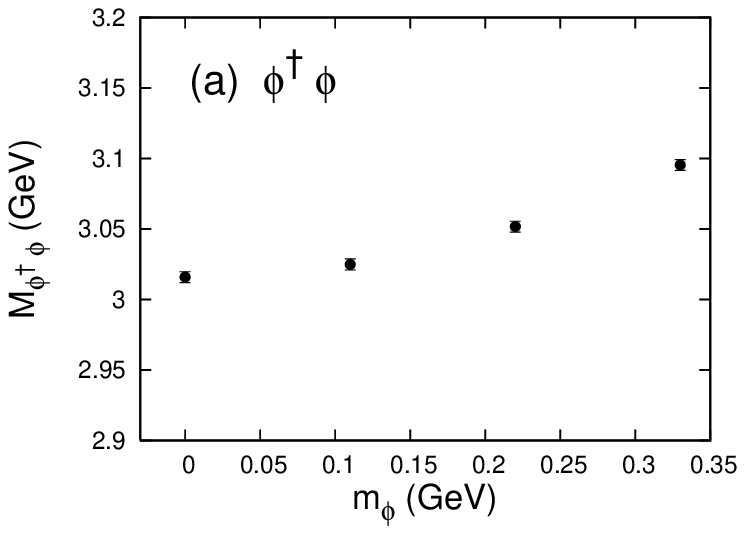} 
\hspace{-0.3cm}
&
\includegraphics[width=4.5cm]{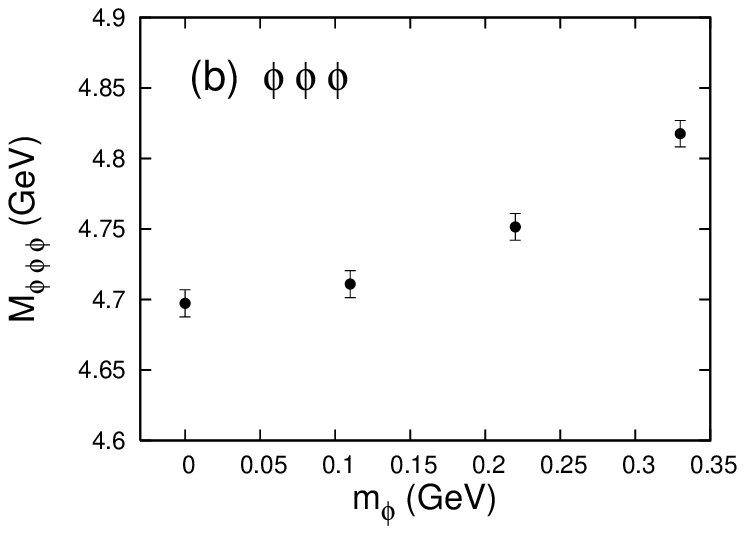}\\
\includegraphics[width=4.5cm]{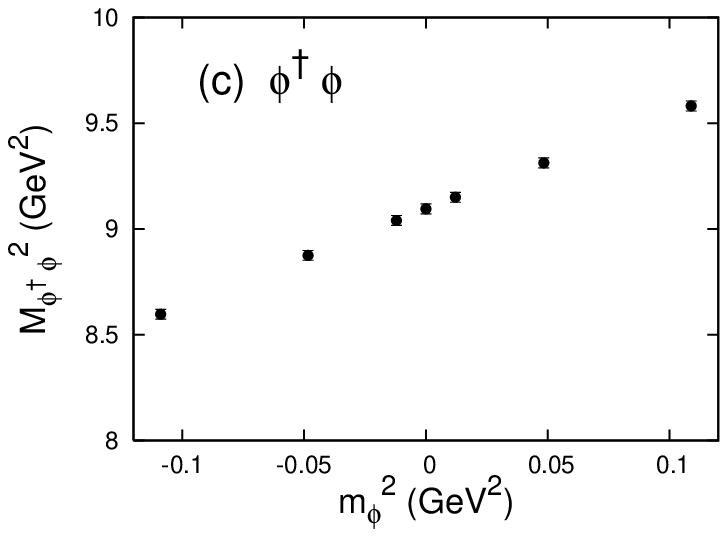} 
\hspace{-0.3cm}
&
\includegraphics[width=4.5cm]{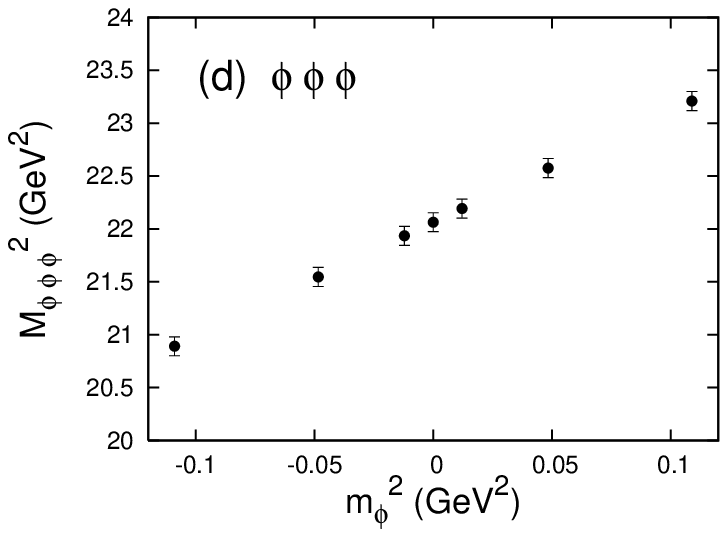}
\end{tabular}
\caption{(a) The scalar-quark meson mass $M_{\phi^\dagger\phi}$ and 
(b) the scalar-quark baryon mass $M_{\phi\phi\phi}$ plotted 
against the bare scalar-quark mass $m_\phi$. 
Even at $m_\phi=0$, 
a large masses of about 3GeV and 4.7GeV are observed 
for the scalar-quark meson and the scalar-quark baryon, respectively.
(c) The scalar-meson mass squared $M_{\phi^\dagger\phi}^2$ and 
(d) the scalar-quark baryon mass squared $M_{\phi\phi\phi}^2$ plotted 
against $m_\phi^2$ in both cases of $m_\phi^2\ge 0$ and $m_\phi^2<0$. 
Calculations can be performed 
even with $m_\phi^2\le 0$. 
Almost linear $m_\phi^2$-dependence is observed 
for both scalar-quark mesons and scalar-quark baryons.
}
\end{center}
\label{fig4}
\end{figure}


In the quark system, the propagator of quarks at $m_\psi=0$ cannot 
be calculated practically in lattice QCD simulations. 
However, in the scalar-quark system, we can calculate 
the propagator of scalar-quarks even at $m_\phi=0$. 
Figures 4(c) and 4(d) show the scalar-quark meson mass squared 
$M_{\phi^\dagger \phi}^2$ and 
scalar-quark baryon mass squared $M_{\phi\phi\phi}^2$ 
including the results for $m_\phi^2<0$ 
plotted against $m_\phi^2$. 
Table \ref{tab5} shows the lattice data for $m_\phi^2<0$. 
As a remarkable fact, the lattice calculation can be performed 
even with $m_\phi^2<0$, 
because the large quantum correction makes the hadron masses positive 
even with such a ``tachyonic'' bare scalar-quark mass. 

In Figs.~4(c) and 4(d), we find the linear $m_\phi^2$-dependence 
of $M_{\phi^\dagger\phi}^2$ and $M_{\phi\phi\phi}^2$. 
More precisely, we find the approximate relations 
\begin{eqnarray}
M_{\phi^\dagger\phi}^2 \simeq 4m_\phi^2 + {\rm const.} 
\end{eqnarray}
and 
\begin{eqnarray}
M_{\phi\phi\phi}^2 \simeq 9m_\phi^2 + {\rm const.}
\end{eqnarray}
These relations can be naturally explained with 
the quantum correction for the scalar field as 
\begin{eqnarray}
M^2_\phi \simeq m_\phi^2+\Sigma_\phi, 
\label{scalarmass}
\end{eqnarray}
together with 
the ``constituent scalar-quark picture" as 
$M_{\phi^\dagger \phi} \simeq 2M_{\phi}$ 
and $M_{\phi\phi\phi}\simeq 3M_{\phi}$.
Here, $\Sigma_\phi$ denotes the self-energy of 
the scalar-quark field $\phi$, 
and its $m_\phi$-dependence is expected to be rather small 
for the region of $|m_\phi^2| \ll \Sigma_\phi$.
Note that Eq.(\ref{scalarmass}) is the natural relation 
between the renormalized mass and the bare mass for scalar particles. 

Figures 5(a) and 5(b) show $M_{\phi^\dagger \phi}^2$ and $M_{\phi\phi\phi}^2$ 
plotted against the bare scalar-quark mass squared $m_\phi^2$ 
including the negative region as $m_\phi^2\ge -(1.2{\rm GeV})^2$. 
Note again that we can perform the lattice calculation even for 
some negative-value region of $m_\phi^2$ 
due to the large quantum correction, i.e., the large self-energy $\Sigma_\phi$.
The calculation breaks down technically 
at about $m_\phi^2 \simeq -(1.2{\rm GeV})^2$,
where the scalar-quark hadron masses becomes almost zero. 
In the actual lattice calculation, 
the iteration process for obtaining the scalar-quark propagator 
does not converge in the region about $m_\phi^2<-(1.2{\rm GeV})^2$. 

To enforce the validity of the constituent scalar-quark picture, 
we show in Fig.~5(c) the ratio of the scalar-quark baryon mass 
$M_{\phi\phi\phi}$ to the scalar-quark meson mass 
$M_{\phi^\dagger\phi}$, i.e., 
$M_{\phi\phi\phi} / M_{\phi^\dagger\phi}$, 
plotted against the bare scalar-quark mass $m_\phi^2$. 
Except for the large negative region of $m_\phi^2$, 
we find the relation of the constituent scalar-quark picture as 
\begin{eqnarray}
\frac{M_{\phi\phi\phi}}{M_{\phi^\dagger\phi}} \simeq 1.5(=\frac{3}{2}),
\end{eqnarray}
where 3/2 is the ratio of the ``constituent number" of 
the scalar-quarks in the new-type hadrons. 



\begin{table*}[ht]
\caption{The mass of scalar-quark hadrons in the region of 
$m_\phi^2<0$. The unit of mass is GeV. 
Notations are the same as those in Table \ref{tab4}.}
\label{tab5}
\begin{tabular}{ccc}
\hline\hline
$m_{\phi}^2$ & \ $M_{\phi^\dagger \phi}$ ($\chi^2/N_{\rm df}$, [fit range]) 
& $ \ M_{\phi\phi\phi}$ ($\chi^2/N_{\rm df}$, [fit range]) \\
\hline
$-(0.11)^2$ & \ 3.007$\pm$0.0039 (0.873, [7-14]) 
& \ 4.684$\pm$0.0096 (0.400, [7-15])\\
$-(0.22)^2$ & \ 2.979$\pm$0.0039 (0.864, [7-14]) 
& \ 4.642$\pm$0.0097 (0.390, [7-15])\\
$-(0.33)^2$ & \ 2.932$\pm$0.0040 (0.847, [7-14]) 
& \ 4.571$\pm$0.0098 (0.372, [7-15])\\
$-(0.55)^2$ & \ 2.780$\pm$0.0041 (0.787, [7-14]) 
& \ 4.328$\pm$0.0102 (0.327, [7-15])\\
$-(0.77)^2$ & \ 2.492$\pm$0.0043 (0.689, [7-14]) 
& \ 3.908$\pm$0.0110 (0.308, [7-15])\\
$-(0.99)^2$ & \ 2.006$\pm$0.0047 (0.622, [7-14]) 
& \ 3.188$\pm$0.0130 (0.548, [7-15])\\
$-(1.10)^2$ & \ 1.586$\pm$0.0051 (0.525, [7-14]) 
& \ 2.587$\pm$0.0159 (0.959, [7-15])\\
$-(1.15)^2$ & \ 1.289$\pm$0.0058 (0.456, [7-14]) 
& \ 2.179$\pm$0.0206 (1.319, [7-15])\\
$-(1.18)^2$ & \ 1.023$\pm$0.0072 (1.359, [7-14]) 
& \ 1.828$\pm$0.0293 (1.683, [7-15])\\
$-(1.20)^2$ & \ 0.717$\pm$0.0115 (0.499, [7-12]) 
& \ 1.324$\pm$0.1144 (2.439, [11-15])\\
\hline\hline
\end{tabular}
\end{table*}

\begin{figure}[h]
\begin{center}
\begin{tabular}{cc}
\includegraphics[width=4.5cm]{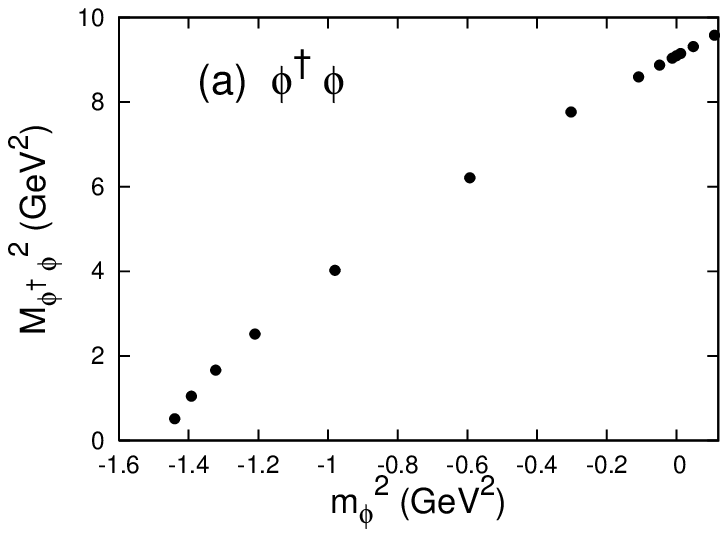} 
\hspace{-0.3cm}
&
\includegraphics[width=4.5cm]{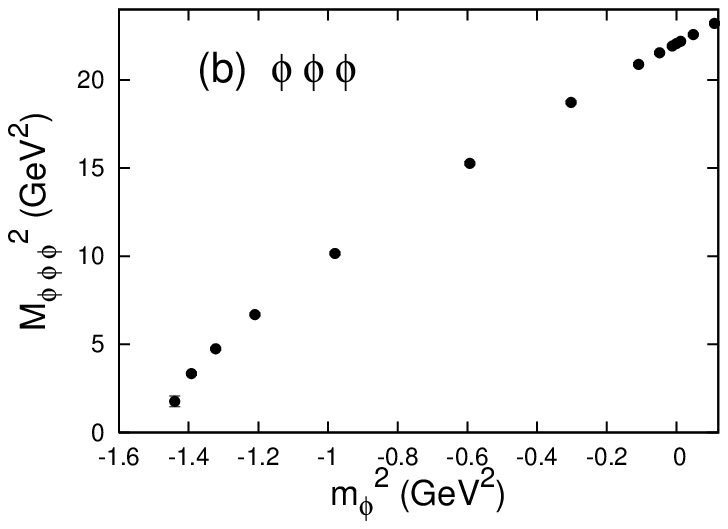}\\
\multicolumn{2}{c}{\includegraphics[width=4.5cm]{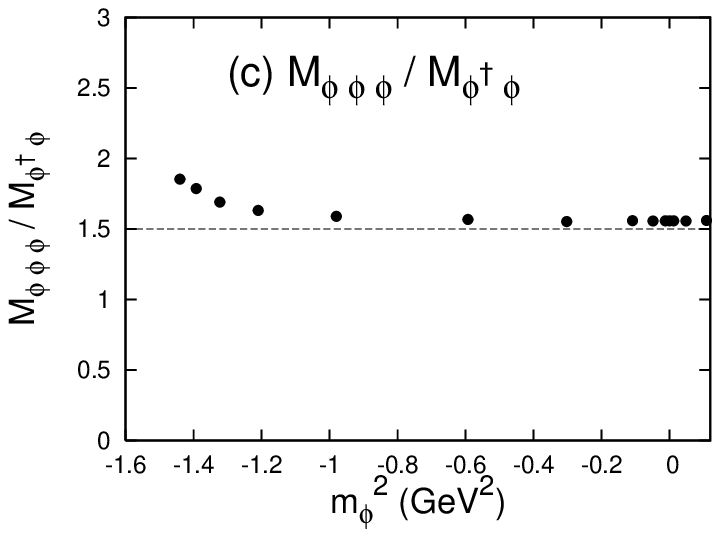}}

\end{tabular}
\caption{
The lattice results for scalar-quark hadrons for 
the negative region of bare scalar-quark mass squared $m_\phi^2$: 
(a) scalar-quark meson mass squared $M_{\phi^\dagger\phi}^2$ v.s. $m_\phi^2$,
(b) scalar-quark baryon mass squared $M_{\phi\phi\phi}^2$ v.s. $m_\phi^2$, 
and (c) the ratio $M_{\phi\phi\phi} / M_{\phi^\dagger\phi}$ v.s. $m_\phi^2$.
Both the scalar-quark meson mass and 
the scalar-quark baryon mass become almost zero 
at about $m_\phi^2 \simeq -(1.2{\rm GeV})^2$. 
Except for the large negative region of $m_\phi^2$, 
one observes the relation of the ``constituent scalar-quark picture" 
as $M_{\phi\phi\phi} / M_{\phi^\dagger\phi} \simeq 1.5$ 
denoted by the dotted line. 
}
\end{center}
\label{fig5}
\end{figure}


\section{Results for chimera hadrons}
\label{sec5}

Next, we show lattice QCD results for the mass of chimera hadrons: 
chimera mesons $\phi^\dagger\psi$ and chimera baryons, $\psi\psi\phi$ and 
$\phi\phi\psi$. 
Figures 6, 7 and 8 show the mass of chimera mesons $\phi^\dagger \psi$ and 
chimera baryons, $\psi\psi\phi$ and $\phi\phi\psi$, 
plotted against the pion mass $M_\pi^2$. 
The panels (a), (b), (c) and (d) of each figure are the masses of each hadron 
at $m_\phi=0$, 110, 220 and 330MeV, respectively, plotted against $M_\pi^2$. 
The data at $M_\pi^2=0$ is obtained from linear extrapolation 
and the solid lines denote the 
best-fit linear functions of the data. 
In panel (e) of each figure, the central values of panels (a) (b) (c) (d) 
are plotted all together.
Different symbols of panel (e) of each figure correspond to 
the different bare scalar-quark mass $m_\phi$ 
(circle: $m_\phi$=0, square: $m_\phi$=0.11GeV, triangle: $m_\phi$=0.22GeV, 
diamond: $m_\phi$=0.33GeV).  
Table \ref{tab6} is the summary of the mass of chimera hadrons. 

\begin{figure}
\begin{center}
\begin{tabular}{cc}
\includegraphics[width=4cm]{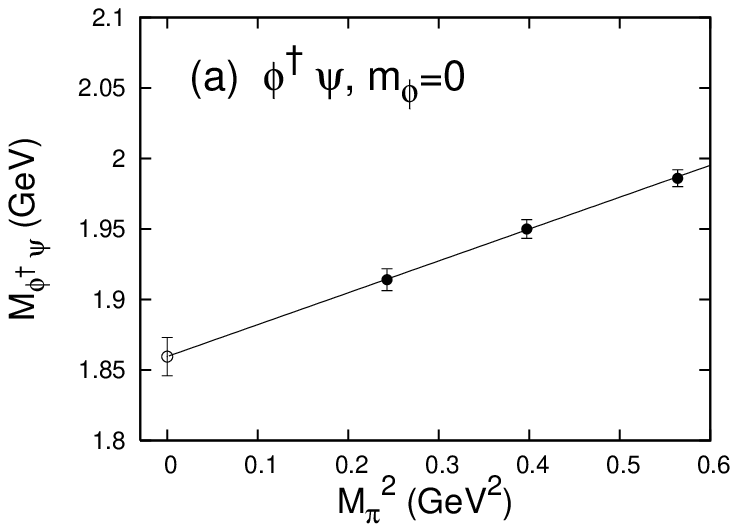} 
\hspace{-0.4cm}
&
\includegraphics[width=4cm]{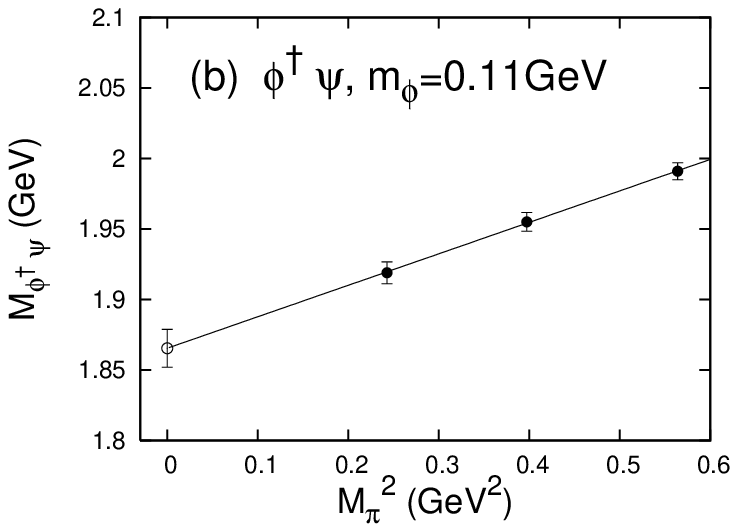} \\
\includegraphics[width=4cm]{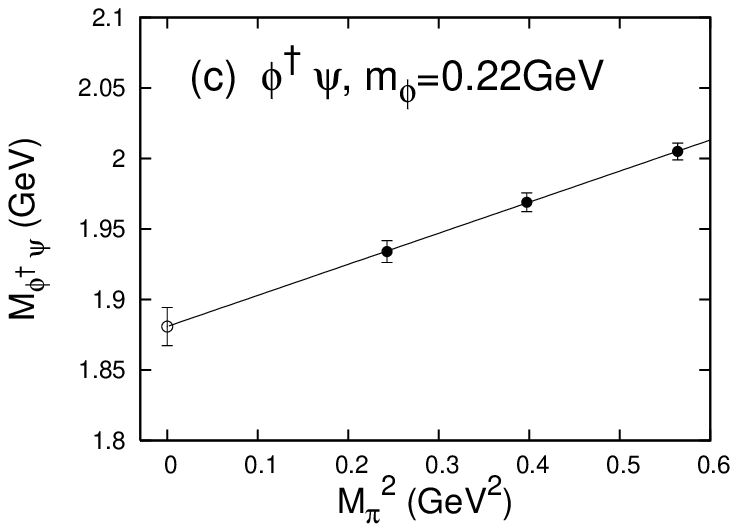} 
\hspace{-0.4cm} 
&
\includegraphics[width=4cm]{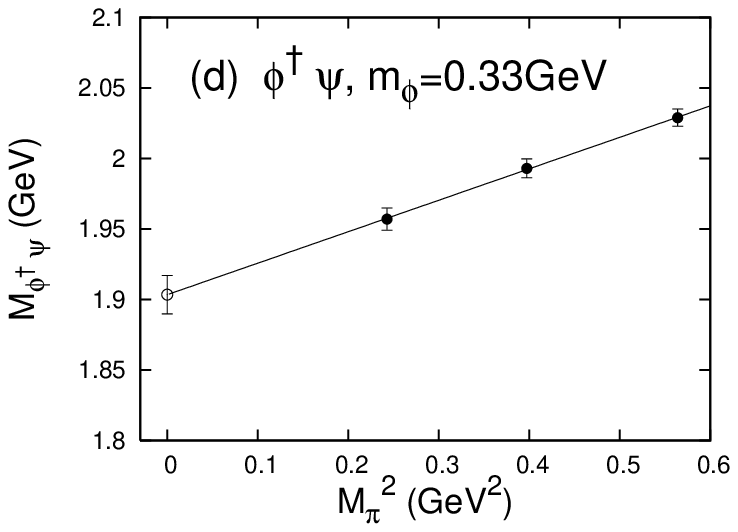} \\
\multicolumn{2}{c}{\includegraphics[width=5.5cm]{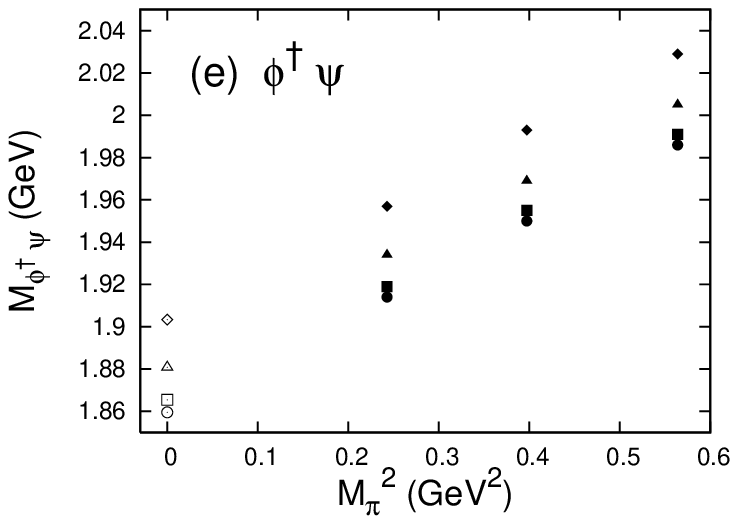}}
\end{tabular}
\caption{The mass of chimera mesons, $M_{\phi^\dagger \psi}$, 
plotted against pion mass squared $M_\pi^2$ in the unit of GeV$^2$. 
(a), (b), (c) and (d) are $M_{\phi^\dagger \psi}$ at $m_\phi=0$, 110, 220 and 330MeV, respectively, 
plotted against $M_\pi^2$. 
The data at $M_\pi^2=0$ is obtained from linear extrapolation, and the solid line denote the 
best-fit linear function of the data. 
In (e), the central values of (a), (b), (c) and (d) are plotted all together. 
Different symbols correspond to the different bare scalar-quark mass $m_\phi$ 
(circle: $m_\phi$=0, square: $m_\phi$=0.11GeV, triangle: $m_\phi$=0.22GeV, diamond: $m_\phi$=0.33GeV).}
\label{fig6}
\end{center}
\end{figure}


\begin{figure}
\begin{center}
\begin{tabular}{cc}
\includegraphics[width=4cm]{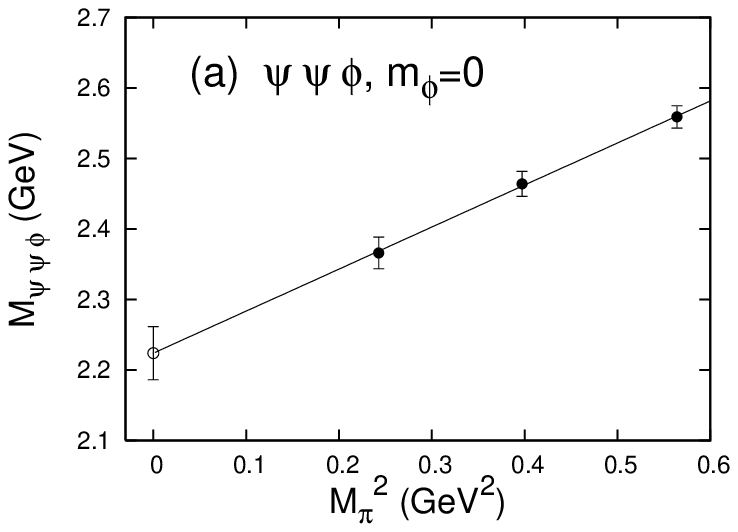} 
\hspace{-0.4cm} 
&
\includegraphics[width=4cm]{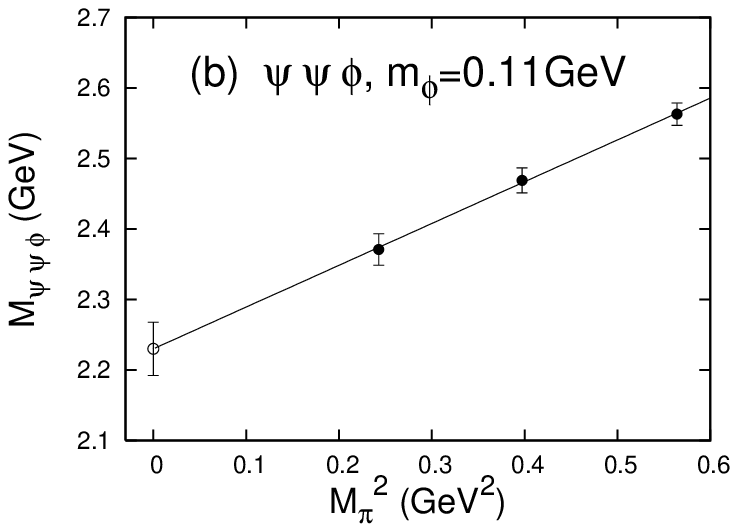} \\
\includegraphics[width=4cm]{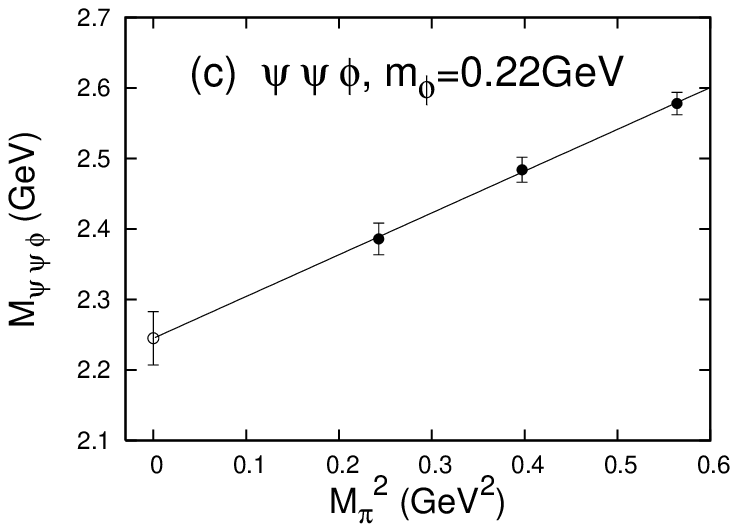} 
\hspace{-0.4cm}
&
\includegraphics[width=4cm]{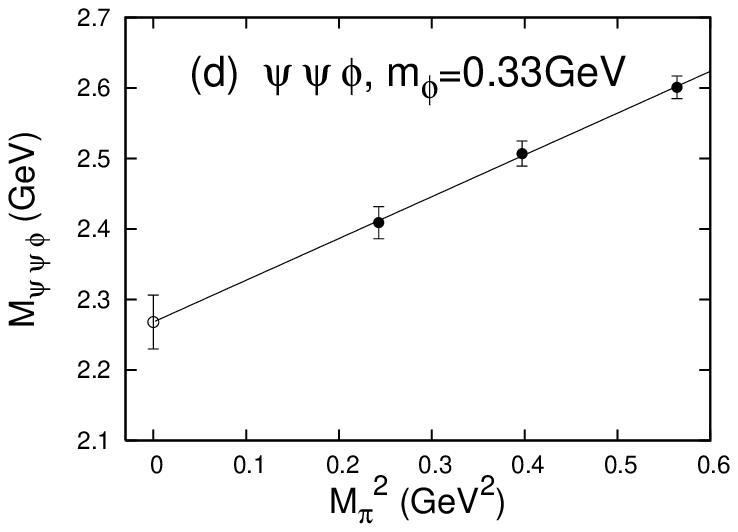} 
\\
\multicolumn{2}{c}{\includegraphics[width=5.5cm]{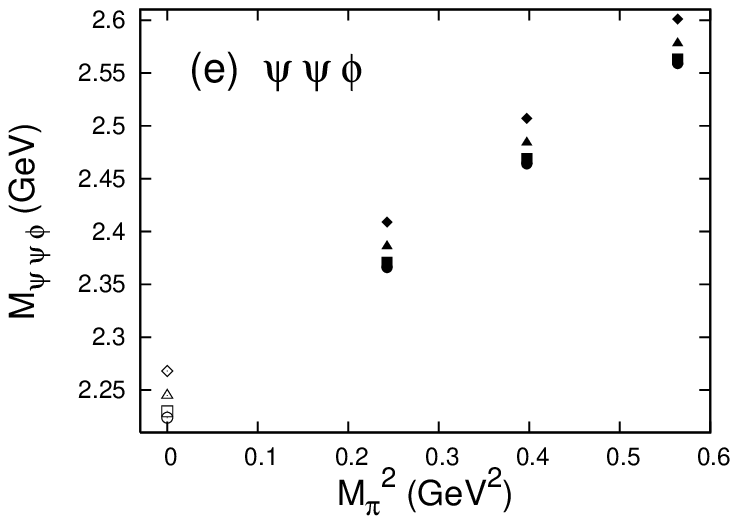}}
\end{tabular}
\caption{The mass of chimera baryons, $M_{\psi\psi\phi}$, 
plotted against pion mass squared $M_\pi^2$ in the unit of GeV$^2$.
The notation, unit and notices are same as that of Fig.~6}
\label{fig7}
\end{center}
\end{figure}


\begin{figure}
\begin{center}
\begin{tabular}{cc}
\includegraphics[width=4cm]{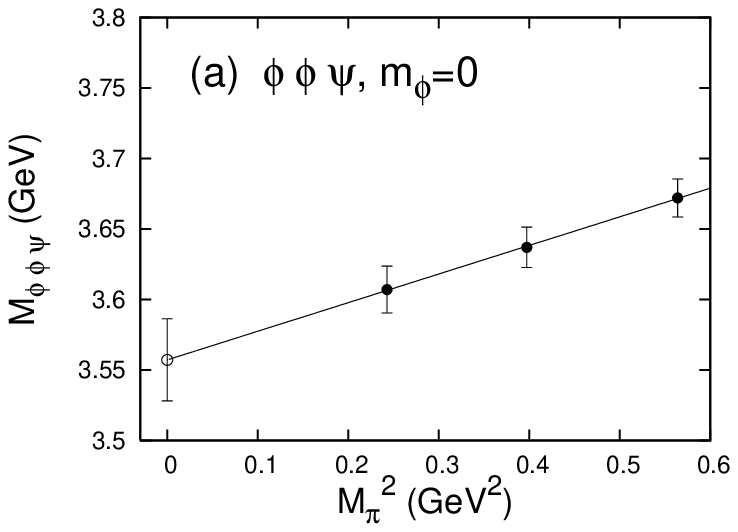}
\hspace{-0.4cm}
&
\includegraphics[width=4cm]{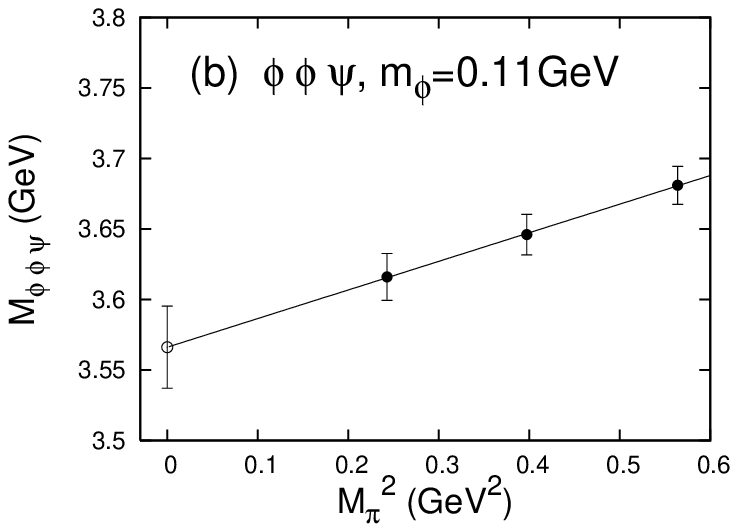} \\
\includegraphics[width=4cm]{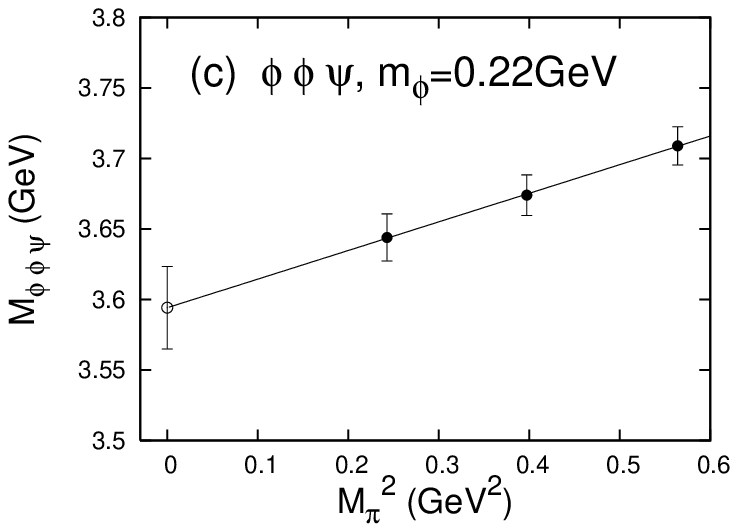} 
\hspace{-0.4cm}
&
\includegraphics[width=4cm]{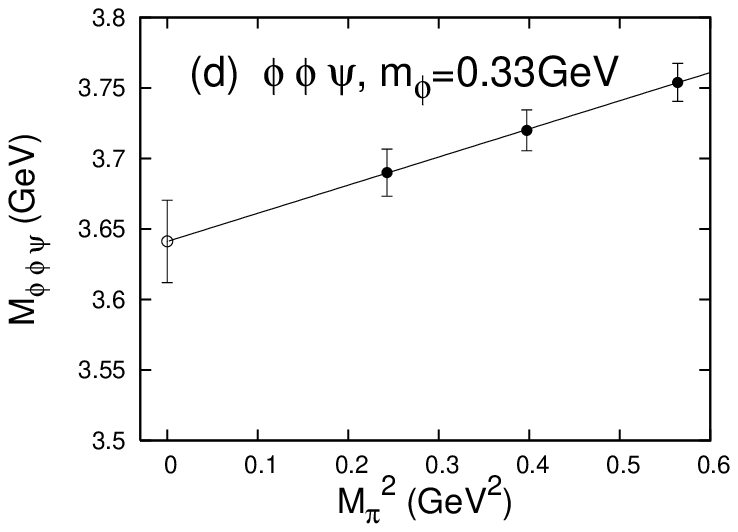} \\
\multicolumn{2}{c}{\includegraphics[width=5.5cm]{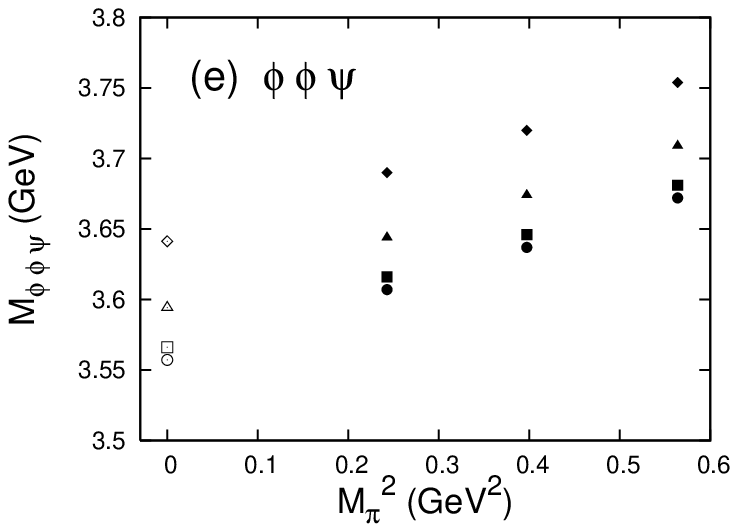}}
\end{tabular}
\caption{The mass of chimera baryons, $M_{\phi\phi\psi}$, 
plotted against pion mass squared $M_\pi^2$ in the unit of GeV$^2$.
The notation, unit and notices are same as that of Fig.~6. }
\label{fig8}
\end{center}
\end{figure}


In Fig.~6, the large mass of chimera mesons $\phi^\dagger \psi$ can be seen. 
The extrapolated value of $M_{\phi^\dagger \psi}$ at $M_\pi^2=0$ 
and $m_\phi=0$ is 1.86GeV. 
Similar to scalar-quark hadron masses, $M_{\phi^\dagger\psi}$ is 
very large compared to ordinary low-lying mesons composed of light quarks. 
Also, we find ``constituent scalar-quark/quark picture", i.e., 
$M_{\phi^\dagger\psi}\simeq M_\phi+M_\psi$ for $M_\phi \simeq $1.5GeV and $M_\psi\simeq$ 400MeV. 
The constituent scalar-quark mass $M_\phi\simeq$1.5GeV is 
consistent with that estimated from scalar-quark hadron masses. 
Concerning the dependence of $m_\phi$ and $M_\pi$, $m_\phi$ dependence is 
rather weak compared to $M_\pi$. In fact, the mass difference between 
$M_{\phi^\dagger\psi}(m_\phi=0, M_\pi^2=0)$ and 
$M_{\phi^\dagger\psi}(m_\phi=0.33{\rm GeV}, M_\pi^2=0)$ is only 43MeV, 
while the difference of $m_\phi$ is 330MeV 
(we refer the mass $M$ at $m_\phi=\alpha$ and $M_\pi=\beta$ to 
$M(m_\phi=\alpha, M_\pi^2=\beta^2)$). 
On the other hand, the mass difference between $M_{\phi^\dagger\psi}(m_\phi=0, 
M_\pi^2=0)$ and $M_{\phi^\dagger\psi}(m_\phi=0, M_\pi^2=(0.75{\rm GeV})^2)$ 
is about 120MeV. This value is the same order of the difference of $m_\psi$, $\delta m_\psi\simeq 190{\rm MeV}$, 
estimated from the relation $2m_\psi a=1/\kappa-1/\kappa_c$.

\begin{table*}[t]
\caption{The mass of chimera hadrons 
(bound states of scalar-quarks $\phi$ and quarks $\psi$) 
in term of the bare scalar-quark mass $m_{\phi}$ and hopping parameter 
$\kappa$ at $a^{-1} \simeq 1.1{\rm GeV}$.
$M_{\phi^\dagger\psi}$, $M_{\psi\psi\phi}$ and $M_{\phi\phi\psi}$
denote the mass of chimera mesons $\phi^\dagger\psi$ and 
chimera baryons ($\psi\psi\phi$, $\phi\phi\psi$), respectively.
The unit is GeV. $\chi^2/N_{\rm df}$ and fit ranges are also listed. 
The values at $\kappa=\kappa_c$, i.e., $M_\pi=0$, are obtained from linear extrapolation. 
Error estimate is done by the jackknife method.}
\label{tab6}
\begin{tabular}{ccccc}
\hline\hline
\ $m_{\phi}$&$\kappa$ & $ \ M_{\phi^\dagger\psi}$ ($\chi^2/N_{\rm df}$, [fit range]) & $ \ M_{\psi\psi\phi}$ ($\chi^2/N_{\rm df}$, [fit range]) & $ \ M_{\phi\phi\psi}$ ($\chi^2/N_{\rm df}$, [fit range])\\
\hline
\multicolumn{1}{c}{0.00}&\multicolumn{1}{c}{$\kappa_c$}& \multicolumn{1}{l}{\ 1.862$\pm$0.013} & \multicolumn{1}{l}{\ 2.230$\pm$0.037} & \multicolumn{1}{l}{\ 3.558$\pm$0.029} \\
&0.1650& \ 1.914$\pm 0.008$ (0.485, [7-12]) & \ 2.366$\pm 0.022$ (0.041, [8-11]) & \ 3.607$\pm 0.017$ (0.142, [7-12])\\
&0.1625& \ 1.950$\pm 0.007$ (0.437, [7-12]) & \ 2.464$\pm 0.018$ (0.173, [8-11]) & \ 3.637$\pm 0.014$ (0.220, [7-12])\\
&0.1600& \ 1.986$\pm 0.006$ (0.460, [7-12]) & \ 2.558$\pm 0.016$ (0.615, [8-11]) & \ 3.672$\pm 0.013$ (0.383, [7-12])\\
\multicolumn{1}{c}{0.11}&\multicolumn{1}{c}{$\kappa_c$}& \multicolumn{1}{l}{\ 1.867$\pm$0.013} & \multicolumn{1}{l}{\ 2.235$\pm$0.037} & \multicolumn{1}{l}{\ 3.568$\pm$0.029} \\
&0.1650& \ 1.929$\pm 0.008$ (0.483, [7-12]) & \ 2.371$\pm 0.022$ (0.043, [8-11]) & \ 3.616$\pm 0.017$ (0.143, [7-12])\\
&0.1625& \ 1.955$\pm 0.007$ (0.436, [7-12]) & \ 2.470$\pm 0.018$ (0.169, [8-11]) & \ 3.646$\pm 0.014$ (0.220, [7-12])\\
&0.1600& \ 1.991$\pm 0.006$ (0.459, [7-12]) & \ 2.564$\pm 0.016$ (0.608, [8-11]) & \ 3.681$\pm 0.013$ (0.383, [7-12])\\
\multicolumn{1}{c}{0.22}&\multicolumn{1}{c}{$\kappa_c$}& \multicolumn{1}{l}{\ 1.883$\pm$0.013} & \multicolumn{1}{l}{\ 2.250$\pm$0.037} & \multicolumn{1}{l}{\ 3.596$\pm$0.029} \\
&0.1650& \ 1.934$\pm 0.008$ (0.478, [7-12]) & \ 2.386$\pm 0.022$ (0.051, [8-11]) & \ 3.644$\pm 0.017$ (0.145, [7-12])\\
&0.1625& \ 1.969$\pm 0.007$ (0.432, [7-12]) & \ 2.484$\pm 0.018$ (0.155, [8-11])& \ 3.674$\pm 0.014$ (0.219, [7-12])\\
&0.1600& \ 2.005$\pm 0.006$ (0.454, [7-12]) & \ 2.578$\pm 0.016$ (0.587, [8-11]) & \ 3.709$\pm 0.013$ (0.382, [7-12])\\
\multicolumn{1}{c}{0.33}&\multicolumn{1}{c}{$\kappa_c$}& \multicolumn{1}{l}{\ 1.905$\pm$0.013} & \multicolumn{1}{l}{\ 2.273$\pm$0.038} & \multicolumn{1}{l}{\ 3.643$\pm$0.029} \\
&0.1650& \ 1.957$\pm 0.008$ (0.470, [7-12]) & \ 2.409$\pm 0.023$ (0.064, [8-11]) & \ 3.690$\pm 0.017$ (0.150, [7-12])\\
&0.1625& \ 1.993$\pm 0.007$ (0.425, [7-12]) & \ 2.507$\pm 0.018$ (0.135, [8-11]) & \ 3.720$\pm 0.014$ (0.216, [7-12])\\
&0.1600& \ 2.029$\pm 0.006$ (0.454, [7-12]) & \ 2.601$\pm 0.016$ (0.554, [8-11]) & \ 3.754$\pm 0.013$ (0.381, [7-12])\\
\hline\hline

\end{tabular}
\end{table*}


As for the chimera baryon $\psi\psi\phi$, the same tendency 
as chimera mesons $\phi^\dagger\psi$ is observed in Figs.~6 and 7 
in terms of $m_\phi$ and $M_\pi$ dependence. 
We find also the ``constituent scalar-quark/quark picture": 
$M_{\psi\psi\phi} \simeq M_\phi+2M_\psi$ for $M_\phi\simeq 1.5{\rm GeV}$ 
and $M_\psi\simeq 400{\rm MeV}$. 
$m_\phi$-dependence of $M_{\psi\psi\phi}$ is weak and is almost 
the same as that of chimera meson.
For $M_\pi$-dependence, 
the difference between $M_{\psi\psi\phi}(m_\phi=0, 
M_\pi^2=0)$ and $M_{\psi\psi\phi}(m_\phi=0, 
M_\pi^2=(0.75{\rm GeV})^2)$ is about 320MeV. 
This is about the same value of 2$\delta m_\psi\simeq 380{\rm MeV}$. 
The factor 2 is from the fact that the chimera baryon $\psi\psi\phi$ includes 
two $\psi$'s.

For the chimera baryon $\phi\phi\psi$, we can see that 
it has the same tendency as chimera mesons 
$\phi^\dagger \psi$ and chimera baryons $\psi\psi\phi$ 
in terms of $m_\phi$ and $M_\pi$ dependence 
from Figs.~6, 7 and 8. 
Constituent scalar-quark/quark picture, 
$M_{\phi\phi\psi} \simeq 2M_\phi+M_\psi$ for $M_\phi\simeq 1.6{\rm GeV}$ and $M_\psi\simeq 400{\rm MeV}$, is satisfied. 
The magnitude of $m_\phi$-dependence of $M_{\phi\phi\psi}$  is about 
twice larger than that of chimera mesons, 
because the chimera baryon $\phi\phi\psi$ has two $\phi$'s. 
$m_\psi$-dependence is almost the same as that of chimera mesons.

The small $m_\phi$-dependence of chimera hadrons can be explained 
by the following simple theoretical consideration. 
As was already mentioned in Sec. \ref{sec4}, 
we find the relation among $M_\phi$, $m_\phi$ and $\Sigma_\phi$ as
$M_\phi^2\simeq m_\phi^2+\Sigma_\phi$ in Eq.(\ref{scalarmass}), 
where the self-energy $\Sigma_\phi$ has only weak $m_\phi$-dependence 
and is almost constant. 
On the other hand, the mass of the chimera hadron 
including $n_\phi \phi$'s and $n_\psi \psi$'s, 
$M_{n_\phi\phi+n_\psi\psi}$, is approximately described by 
\begin{eqnarray}
M_{n_\phi\phi+n_\psi\psi} \simeq n_\phi M_\phi+n_\psi M_\psi.
\label{chimeramass}
\end{eqnarray}
Using Eqs.~(\ref{scalarmass}) and (\ref{chimeramass}), 
we obtain for $|m_\phi^2| \ll \Sigma_\phi$ 
the relation between $M_{n_\phi\phi+n_\psi\psi}$ and $m_\phi^2$ as 
\begin{eqnarray}
M_{n_\phi\phi+n_\psi\psi} &\simeq& 
(n_\phi\sqrt{\Sigma_\phi}+n_\psi M_\psi) 
+\frac{n_\phi}{2}\frac{m_\phi^2}{\sqrt{\Sigma_\phi}} \nonumber \\
&\simeq& Km_\phi^2+(m_\phi\hbox{-}{\rm independent~part})
\label{massrel}
\end{eqnarray}
up to the first order of the expansion by $m_\phi^2/\Sigma_\phi$. 
In our calculation range of $|m_\phi^2| \le (0.33{\rm GeV})^2$, 
$|m_\phi^2|$ is much smaller than $\Sigma_\phi \simeq$ (1.5GeV)$^2$, 
i.e., $|m_\phi^2|/\Sigma_\phi \simeq 0.048 (\ll 1)$, 
and hence this expansion is expected to be rather good. 
The slope parameter $K$ is theoretically conjectured as 
$K = n_\phi/(2\sqrt{\Sigma_\phi})$.
We show in Fig.~9 the chimera meson mass $M_{\phi^\dagger \psi}$ and 
the chimera baryon mass, $M_{\psi\psi\phi}$ and $M_{\phi\phi\psi}$, 
plotted against the bare scalar-quark mass squared $m_\phi^2$ 
at the pion mass $M_\pi \simeq 0.49{\rm GeV}$. 
In the figure, we observe the linear $m_\phi^2$-dependence 
for all the chimera hadrons. 
The slopes of these lines are 0.395GeV$^{-1}$ for $\phi^\dagger\psi$, 0.395GeV$^{-1}$ for $\psi\psi\phi$ and 
0.767GeV$^{-1}$ for $\phi\phi\psi$ from the lattice data. On the other hand, 
the slope $K$ is $n_\phi/(2\sqrt{\Sigma_\phi})$ for the chimera hadron including $n_\phi \phi$'s from the theoretical conjecture of Eq.(\ref{massrel}): 
the theoretical estimate of the slope $K$ is 
0.333GeV$^{-1}$ for $\phi^\dagger\psi$, 
0.333GeV$^{-1}$ for $\psi\psi\phi$, 
and 0.666GeV$^{-1}$ for $\phi\phi\psi$. 
The values of slopes from lattice data approximately agree 
with those from the theoretical estimation. 
Thus, $m_\phi^2$-dependence of the chimera hadron masses 
can be explained with Eq.~(\ref{massrel}), 
and the $m_\phi$-dependence is rather weak 
due to the large self-energy of scalar-quarks, $\Sigma_\phi$.

\begin{figure}[t]
\begin{center}
\begin{tabular}{cc}
\includegraphics[width=4cm]{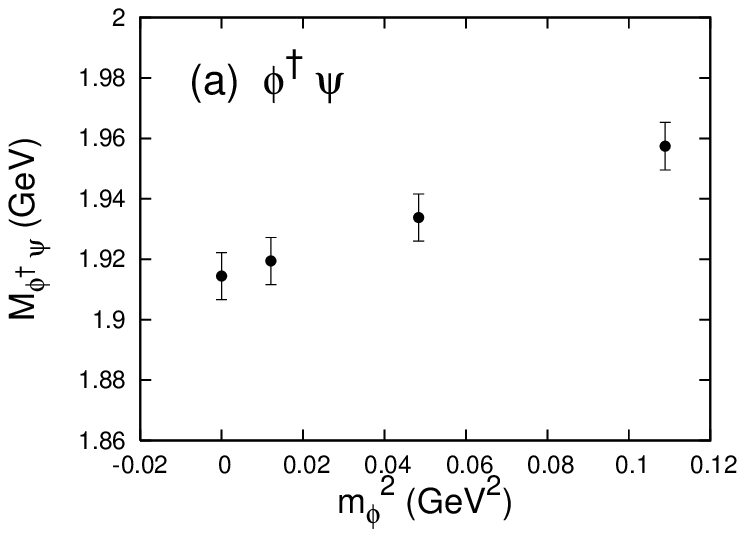} 
\hspace{-0.4cm}
&
\includegraphics[width=4cm]{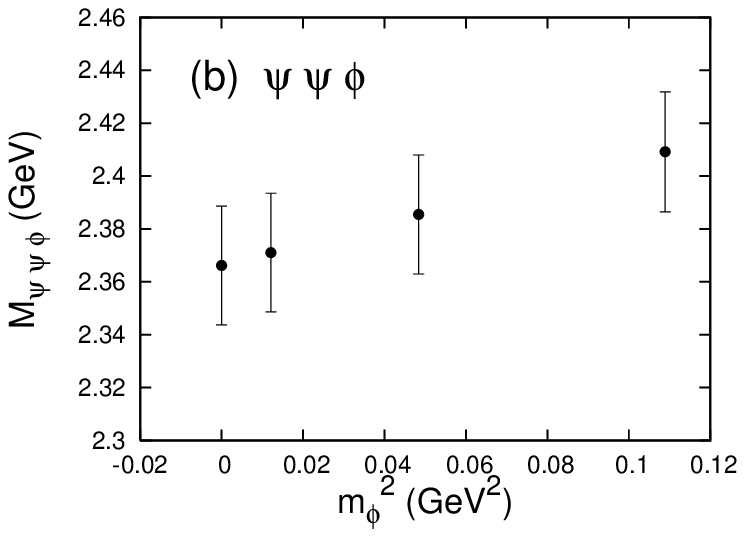} \\
\multicolumn{2}{c}{\includegraphics[width=4cm]{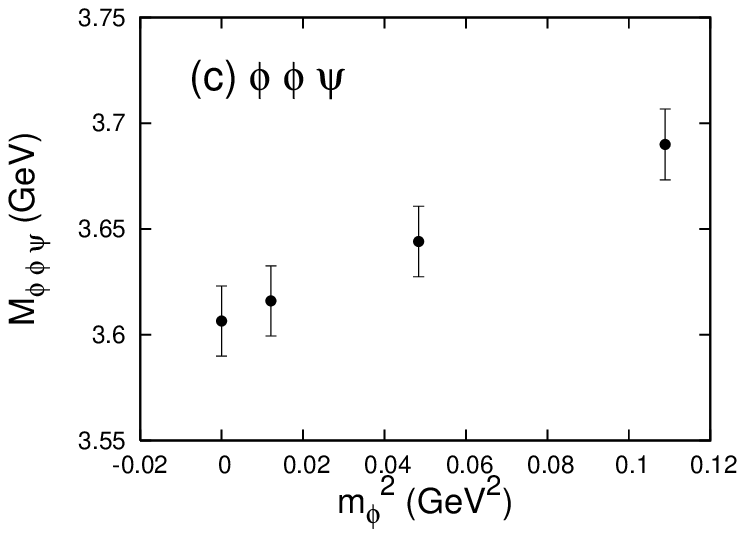}}
\end{tabular}
\vspace{-0.2cm}
\caption{Chimera meson mass $M_{\phi^\dagger \psi}$, chimera baryon mass 
$M_{\psi\psi\phi}$ and chimera baryon mass $M_{\phi\phi\psi}$ plotted 
against bare scalar-quark mass squared $m_\phi^2$ in unit of GeV. 
The pion mass of the figure data is $M_\pi \simeq 0.49{\rm GeV}$. 
Linear dependence of these masses for $m_\phi^2$ can be seen. 
}
\label{fig6}
\end{center}
\end{figure}


Here, we also note the particular similarity between 
chimera mesons $\phi^\dagger\psi$ and $\phi\phi\psi$-type chimera baryons 
in Figs.~6 and 8. 
To see this, 
we investigate the $M_\pi^2$-dependence of chimera mesons $\phi^\dagger\psi$ 
and chimera baryons, $\psi\psi\phi$ and $\phi\phi\psi$. 
The relation between the bare quark mass $m_\psi$, 
the constituent quark mass $M_\psi$ 
and the self-energy for quarks, $\Sigma_\psi$, is described as
\begin{eqnarray}
M_\psi \simeq m_\psi +\Sigma_\psi.
\label{seq}
\end{eqnarray}
The bare quark mass $m_\psi$ is proportional to the pion mass squared $M_\pi^2$ in the small $m_\psi$ region, i.e.,  near the chiral limit, 
from the Gell-Mann-Oaks-Renner relation, 
$f_\pi^2M_\pi^2\simeq -2m_\psi \langle \bar qq \rangle$, i.e.,  
$m_\psi \simeq k M_\pi^2$ with $k \equiv -f_\pi^2/(2\langle\bar qq \rangle)$. 
Inserting Eq.~(\ref{seq}) into Eq.~(\ref{massrel}),  
the mass $M_{n_\phi\phi+n_\psi\psi}$ 
of the chimera hadron including $n_\phi$ $\phi$'s 
and $n_\psi$ $\psi$'s is estimated in terms of $m_\phi^2$ and $M_\pi^2$ as
\begin{eqnarray}
M_{n_\phi\phi+n_\psi\psi}(m_\phi^2, M_\pi^2)
&\simeq& 
(n_\phi\sqrt{\Sigma_\phi} +n_\psi\Sigma_\psi) \nonumber \\
&+&\frac{n_\phi}{2}\frac{m_\phi^2}{\sqrt{\Sigma_\phi}} 
+n_\psi k M_\pi^2.
\end{eqnarray}
We investigate the mass difference, 
\begin{eqnarray}
&&\Delta M_{n_\phi\phi+n_\psi\psi}(m_\phi^2, M_\pi^2) \nonumber \\
&\equiv& M_{n_\phi\phi+n_\psi\psi}(m_\phi^2, M_\pi^2) 
-M_{n_\phi\phi+n_\psi\psi}(m_\phi^2, M_\pi^2=0) \nonumber \\
&\simeq& n_\psi k M_\pi^2,
\end{eqnarray}
for the chimera hadrons including $n_\phi$ $\phi$'s 
and $n_\psi$ $\psi$'s.
Then, we expect the following relation near the chiral limit as 
\begin{eqnarray}
\Delta M_{\phi\phi\psi}\simeq 
\Delta M_{\phi^\dagger\psi}\simeq \Delta M_{\psi\psi\phi}/2 \simeq k M_\pi^2. 
\end{eqnarray}
Figure 10 shows $\Delta M_{\phi^\dagger\psi}$, 
$\Delta M_{\phi\phi\psi}$ and $\Delta M_{\psi\psi\phi}/2$ 
plotted against $M_\pi^2$ at $m_\phi=0, 0.11, 0.22$ and $0.33$GeV. 
We find a clear linear $M_\pi^2$-dependence for these chimera hadrons.
Since the four data with different values of $m_\phi$ almost coincides 
at each case of chimera hadrons, 
there is almost no $m_\phi$-dependence for these mass differences, 
indicating that the quark in chimera hadrons is insensitive 
to the bare scalar-quark mass $m_\phi$.
We find the accurate coincidence between $\Delta M_{\phi^\dagger\psi}$ and
$\Delta M_{\phi\phi\psi}$ within several MeV, and 
some deviation between these two and $\Delta M_{\psi\psi\phi}/2$. 
Therefore, we expect a similar structure between 
the chimera meson $\phi^\dagger\psi$ and the chimera baryon $\phi\phi\psi$, 
while the chimera baryon $\psi\psi\phi$ 
is somewhat different from these two chimera hadrons.

These features can be understood by the following non-relativistic picture. 
Due to the large mass of the scalar-quark, 
the wave-function of $\phi^\dagger$ in the chimera meson $\phi^\dagger \psi$ 
is localized near the center of mass of the system, 
and the wave-function of $\psi$ distributes around the heavy anti-scalar-quark 
as shown in Fig.11(a). 
On the other hand, for chimera baryons $\phi\phi\psi$, 
two $\phi$'s get close in the one-gluon-exchange Coulomb potential 
due to their heavy masses. 
Therefore, two $\phi$'s behave like a point-like ``di-scalar-quark''. 
As a result, the structure of chimera baryons $\phi\phi\psi$ is similar to that of chimera mesons $\phi^\dagger\psi$: 
the wave-function of $\phi\phi$ is localized near the center of mass, 
and that of $\psi$ distributes around $\phi\phi$ as shown in Fig.~11(b). 
The difference between these two systems is only the mass of the heavy object 
($M_\phi$ for $\phi^\dagger \psi$ and 2$M_\phi$ for $\phi\phi\psi$). 
The heavy-light system is mainly characterized by the light object, 
because the heavy object only give the potential and the reduced mass 
in the system is almost the mass of the light object. 
We note that such a similarity between the chimera meson $\phi^\dagger\psi$ 
and the chimera baryon $\phi\phi\psi$ is originated from 
the large quantum correction of bare scalar-quark mass $m_\phi$. 

%

\begin{figure}[t]
\begin{center}
\includegraphics[width=6cm]{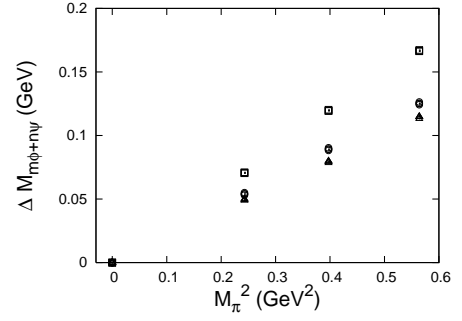} 
\caption{$\Delta M_{\phi^\dagger\psi}\equiv M_{\phi^\dagger\psi}(m_\phi^2, M_\pi^2)-
M_{\phi^\dagger\psi}(m_\phi^2, M_\pi^2=0)$, $\Delta M_{\psi\psi\phi}/2\equiv (M_{\psi\psi\phi}(m_\phi^2, M_\pi^2)-M_{\psi\psi\phi}(m_\phi^2,M_\pi^2=0))/2$ and 
$\Delta M_{\phi\phi\psi}\equiv M_{\phi\phi\psi}(m_\phi^2, M_\pi^2)-M_{\phi\phi\psi}(m_\phi^2,M_\pi^2=0)$ plotted against $M_\pi^2$ at $m_\phi=0, 0.11, 0.22$ and $0.33$GeV. 
Different symbols correspond to the different types of the chimera hadrons 
(triangle: chimera mesons $\phi^\dagger \psi$,  circle: 
chimera baryons $\phi\phi\psi$ and square: chimera baryons $\psi\psi\phi$). 
The four data at $m_\phi=0, 0.11, 0.22$ and $0.33$ 
almost coincide at each case, indicating $m_\phi$-independence 
for these mass differences. 
The difference between $\Delta M_{\phi^\dagger\psi}$
and $\Delta M_{\phi\phi\psi}$ is only several MeV.
}
\end{center}
\label{fig10}
\end{figure}

\begin{figure}[t]
\includegraphics[width=7.5cm]{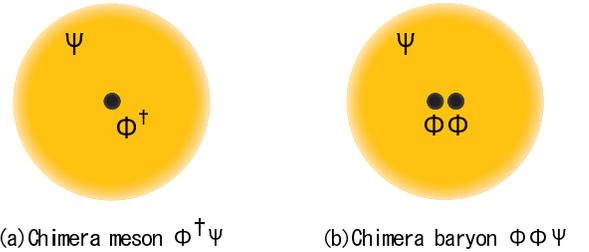}
\caption{Structure of chimera mesons $\phi^\dagger\psi$ and chimera baryons $\phi\phi\psi$. 
The wave-function of the quark $\psi$ in the chimera meson distributes 
around the heavy scalar-quark $\phi^\dagger$. Similar to chimera mesons, 
the wave-function of $\psi$ in chimera baryons distributes 
around the two $\phi$'s localized near the center of mass.}
\label{fig11}

\end{figure}

\section{Conclusion and discussion}
\label{sec6}

We have performed the first study for the light scalar-quarks $\phi$ 
and their color-singlet hadronic states 
using quenched SU(3)$_c$ lattice QCD 
in terms of their mass generation. 
We have investigated the scalar-quark bound states 
(scalar-quark hadrons): scalar-quark mesons 
$\phi^\dagger \phi$ and scalar-quark baryons $\phi\phi\phi$. 
We have also investigated the bound states of scalar-quarks and quarks, 
which we name ``chimera hadrons'': 
chimera mesons $\phi^\dagger \psi$ and chimera baryons, $\psi\psi\phi$ and 
$\phi\phi\psi$. 
For the lattice QCD calculation, we have adopted 
the standard plaquette action for the gluon sector, 
and the Wilson fermion action for the quark sector. 
We have presented the simplest local and gauge-invariant 
lattice action of scalar-quarks. 
We have adopted $\beta\equiv 2N_c/g^2=5.70$, which corresponds to 
the lattice spacing $a\simeq 0.18{\rm fm}=(1.1{\rm GeV})^{-1}$, 
and lattice size $16^3\times 32$ (spatial volume $V=(2.9{\rm fm})^3$). 
For the bare scalar-quark mass, we 
have adopted $m_\phi=0.0, 0.11, 0.22$ and $0.33{\rm GeV}$, 
and for the bare quark mass, 
$\kappa=0.1650, 0.1625$ and $ 0.1600$. 
Fitting the temporal correlator of each new-type hadrons by single exponential function, 
we have extracted the mass of lowest state of the hadrons. 

For the scalar-quark hadrons, 
we have found their large mass. 
The masses of scalar-quark mesons $\phi^\dagger\phi$ and scalar-quark baryons $\phi\phi\phi$ 
at $m_\phi=0$ are about $3$GeV and $4.7$GeV, respectively. 
We have found the constituent scalar-quark picture, 
i.e., $M_{\phi^\dagger\phi}\simeq 2M_{\phi}$ 
and $M_{\phi\phi\phi}\simeq 3M_{\phi}$, where 
$M_{\phi}$ is the constituent scalar-quark mass about 1.5GeV. 
We have also found that, even in the case of $m_\phi^2 \le 0$, 
the calculation can be performed for $m_\phi^2 \ge -(1.2{\rm GeV})^2$. 
$m_\phi$-dependence of scalar-quark hadron masses squared is linear with respect to $m_\phi^2$ 
except for the large negative region of $m_\phi^2$. 
This is natural because the relation between the constituent scalar-quark mass $M_\phi$ 
and bare scalar-quark mass $m_\phi$ is written as $M_\phi^2=m_\phi^2+\Sigma_\phi$. 
For the chimera hadrons, we have also found their large masses: 
$M_{\phi^\dagger\psi}\simeq 1.9{\rm GeV}$, 
$M_{\psi\psi\phi}\simeq 2.2{\rm GeV}$ and 
$M_{\phi\phi\psi}\simeq 3.6{\rm GeV}$, for $m_\phi=0$ and 
$M_\pi$. The constituent scalar-quark/quark picture is also satisfied, i.e., 
$M_{{m}\phi+{n}\psi}\simeq {m}M_\phi+{n}M_\psi$, where constituent scalar-quark mass 
$M_\phi=$1.5-1.6GeV and constituent quark mass $M_\psi=$400MeV. 
$m_\phi$-dependence of chimera-hadron masses is weak 
and the bare quark mass $m_\psi$ seems to 
govern their systems in our calculated range of $m_\phi$ and $m_\psi$. 
The small $m_\phi$-dependence of chimera hadrons can be explained 
by a simple theoretical consideration. 
Due to the large mass generation of $\phi$, there occurs similarity of the 
structure 
between chimera mesons $\phi^\dagger\psi$ and chimera baryons $\phi\phi\psi$. 

The large mass generation of the scalar-quark $\phi$ 
reflects the general argument of large quantum corrections 
for scalar particles. 
As other famous example, the Higgs scalar field suffers from 
large radiative corrections in the Grand Unified Theory (GUT), 
and to realize the low-lying mass of the Higgs scalar inevitably requires 
``fine tuning'' related to the gauge hierarchy problem \cite{CL, hierarchy}.
In fact, while the typical energy scale of the electro-weak unification is 
about $O$($10^2$GeV) and the Higgs scalar mass is to be also $O$($10^2$GeV), 
the GUT scale, i.e., the typical energy scale of the electro-weak 
and strong unification, is conjectured to be extremely large 
as $E_{\rm GUT}\sim 10^{15-16}$GeV.
In the simple GUT scenario, the standard model is regarded as 
an effective theory applicable up to the GUT scale $E_{\rm GUT}$, 
which can be identified as the cutoff scale of the standard model 
as $\Lambda \sim E_{\rm GUT}$. 
For the self-energy of the Higgs scalar, $\Sigma_{\rm H}$, 
there appears a large quantum correction including second-order divergence as  
$\Sigma_{\rm H}\sim \Lambda^2$, with the cutoff parameter $\Lambda$. 
This is because scalar particles do not have symmetry restriction like 
the chiral symmetry for light fermions and the gauge symmetry 
for vector gauge bosons. 
Therefore, it is natural that the Higgs scalar receives 
huge quantum correction of order $E_{\rm GUT}$, 
and, in order to realize the low-lying physical Higgs mass 
of order $10^2$GeV, an extreme fine tuning is necessary 
for the bare mass of the Higgs scalar.
This is the problem of fine tuning for the Higgs scalar. 
In lattice QCD, the self-energy of scalar-quarks $\Sigma_\phi$ 
also receives a large quantum correction including 
the second-order divergence as $\Sigma_\phi \sim \Lambda^2$, 
similar to the Higgs scalar. 
Since the constituent mass $M_\phi$ of the scalar-quark satisfies 
the relation of $M_\phi^2 \simeq m_\phi^2+\Sigma_\phi$ 
with the bare mass $m_\phi$ 
from a general argument of the scalar field, 
the constituent scalar-quark mass $M_\phi$ 
takes a large value even at $m_\phi=0$ 
as $M_\phi \simeq \Sigma^{1/2} \sim \Lambda$.
Note here that the lattice QCD can be regarded 
as a cutoff theory with the cutoff about the lattice spacing 
as $\Lambda \sim a^{-1}$.
Actually, the quantum correction to the scalar-quark mass is found to be 
about $1.5{\rm GeV}$ in this lattice study, which is approximately 
the same order of the cutoff scale $a^{-1}\simeq 1.1{\rm GeV}$. 
(From a finer lattice calculation at $\beta$=6.10 in Appendix, 
the large quantum correction for the scalar-quark mass is conjectured to be 
proportional to the lattice cutoff scale.)
In this way, the large quantum correction to the scalar-quarks 
is naturally explained.

Here, we comment on the continuum limit of the system including scalar-quarks.
As was already shown, the quantum correction for the scalar-quark mass 
is quite large and becomes infinite in the continuum limit, 
which is indicated by a finer lattice calculation at $\beta$=6.10 in Appendix.
To take the continuum limit with keeping the (constituent) 
scalar-quark mass finite, one has to tune the bare scalar-quark mass 
so as to make the related physical quantities like scalar-quark hadron masses 
finite, during the decrease of lattice spacing. 
In the fine-tuned case, the bare scalar-quark mass squared becomes 
negative infinity so as to cancel the infinitely large quantum correction.
In this way, the continuum limit where the (constituent) scalar-quark mass 
is finite can be achieved by tuning the bare mass of 
scalar-quarks in principle. 
However, without such a fine tuning, 
the scalar-quark systems are conjectured to be infinitely massive 
and to lose the physical relevance in the continuum limit, 
because of the large quantum correction for the scalar-quark mass. 
This situation is rather different from that of fermion systems in QCD, 
where such a fine tuning of the bare mass is not needed. 

To include dynamical scalar-quarks is one of the possible future works. 
Since the scalar-quarks are described to be Grassmann-even variables 
like real numbers, 
we can set the scalar-field variable $\phi(x)$ directly on the lattice 
using real numbers, 
and perform the actual Monte Carlo calculation including dynamical scalar-quark
using the standard heat-bath algorithm.
This point is essentially different from the case of fermions:
since fermions are described by Grassmann-odd variables, 
to include the dynamical effect of fermions in the Monte Carlo calculation, 
a huge calculation of fermionic determinant is needed.
In contrast, to include the dynamical scalar-quarks, 
we need not to perform the huge calculation of functional determinant of 
the scalar field.  
However, the dynamical effect of such a ``heavy" scalar-quark 
is expected to be small in the actual calculation, considering the 
large constituent scalar-quark mass generation as $M_\phi \simeq$ 1.5-1.6GeV.

In terms of the diquark picture, we find that 
the point-like diquarks interacting with gauge fields have 
large mass about 1.5GeV at the cutoff $a^{-1}\simeq 1$GeV 
which is about the hadronic scale. 
The diquark mass of about 1.5GeV is too heavy 
to use in effective models of hadrons. 
Therefore, 
the lattice result indicates that the simple modeling 
which treats the diquark as a local scalar field at the scale of 
$a^{-1} \sim 1{\rm GeV}$ in QCD is rather dangerous. 
If we treat the diquark as a point-like object, 
we must do the subtle tuning of the diquark model, 
and the model cannot be so simple one like QCD. 

In this way, even without chiral symmetry breaking, 
large dynamical mass generation occurs for the scalar-quark systems 
as scalar-quark/chimera hadrons.
Together with the large glueball mass ($>$ 1.5GeV) and 
large difference ($\sim$ 400MeV) between current and constituent charm-quark masses, 
this type of mass generation would be generally occurred in the strong interaction, 
and therefore we conjecture that {\it all colored particles generally acquire 
a large effective mass due to dressed gluon effects} as shown in Fig.~12.

\begin{figure}[t]
\includegraphics[width=8cm]{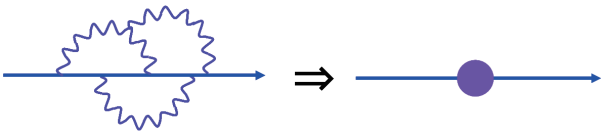}
\caption{Schematic figure for dynamical mass generation of colored particles. 
Even without chiral symmetry breaking, colored particles generally acquire 
a large effective mass due to dressed gluons.}
\label{fig10}
\end{figure}

\begin{acknowledgements}
\vspace{-0.3cm}

{H.~S. is supported in part by the Grant for Scientific Research 
[(C) No.16540236, (C) No.19540287] 
from the Ministry of Education, Culture, Sports, Science and Technology (MEXT) of Japan. 
H.~I. and T.T.~T. are supported by the Japan Society for the Promotion of Science for Young Scientists. 
This work is supported by the Grant-in-Aid for the 21st Century COE 
``Center for Diversity and Universality in Physics" from the MEXT of Japan.
Our lattice QCD calculations have been performed on 
NEC-SX5 and NEC-SX8 at Osaka University. }
\end{acknowledgements}

\appendix
\section{Scalar-quark meson mass at $\beta=6.10$}
\label{appendix}

In this appendix, regarding the scalar-quark as a point-like object, 
we investigate the scalar-quark meson mass 
$M_{\phi^\dagger\phi}$ with a finer lattice at $\beta=6.10$ 
to check the approximate relation that 
the large self-energy for scalar-quarks 
is almost proportional to the cutoff scale as 
$\sqrt{\Sigma_\phi} \propto a^{-1}$. 
The lattice spacing $a$ at $\beta=6.10$ is about twice smaller 
($a\simeq 0.086{\rm fm}\simeq (2.3{\rm GeV})^{-1}$
\cite{DIOS04}) than that at $\beta=5.70$ 
($a\simeq 0.18{\rm fm} \simeq (1.1{\rm GeV})^{-1}$), 
so that the ultraviolet cutoff at $\beta=6.10$ 
is about twice larger than that at $\beta$=5.70. 
We summarize the parameters of the lattice calculation in Table \ref{tab7}. 

\begin{table}[ht]
\caption{Parameters and relevant quantities of lattice QCD calculations 
at $\beta=6.10$. }
\label{tab7}
\begin{tabular}{cccccc}
\hline\hline
$\beta$ &Lattice size& $a^{-1}$& $N_{\rm conf}$ &
bare mass $m_{\phi}$ [$a^{-1}$]\\ 
\hline
$6.10$ & $16^3\times 24$&  2.3 GeV & 100 
& 0.0, 0.25, 0.5, 0.75, 1.0 \\
\hline\hline
\end{tabular}
\end{table}

\begin{table}[h]
\caption{
Scalar-quark meson mass $M_{\phi^\dagger\phi}$ 
in term of the bare scalar-quark mass $m_{\phi}$ 
at $a^{-1} \simeq 2.3{\rm GeV}$.
$\chi^2$ over degrees of freedom $N_{\rm df}$ and 
fit ranges are also listed. 
The error estimate is done by the jackknife method.}
\begin{tabular}{ccc}
\hline\hline
$m_{\phi}$ [$a^{-1}$]& $ \ M_{\phi^\dagger \phi}$ [GeV] \ ($\chi^2/N_{\rm df}$, [fit range]) \\
\hline
0   & \ 5.525$\pm$0.0071 (2.643, [8-10]) \\
0.5 & \ 6.030$\pm$0.0068 (1.495, [9-11]) \\
1.0 & \ 7.244$\pm$0.0070 (0.067, [9-11]) \\
1.5 & \ 8.664$\pm$0.0074 (0.248, [9-11]) \\
2.0 & \ 10.043$\pm$0.0074 (3.212, [9-11]) \\
\hline\hline
\end{tabular}
\label{tab8}
\end{table}

\begin{figure*}[ht]
\begin{center}
\begin{tabular}{cc}
\includegraphics[width=5.5cm]{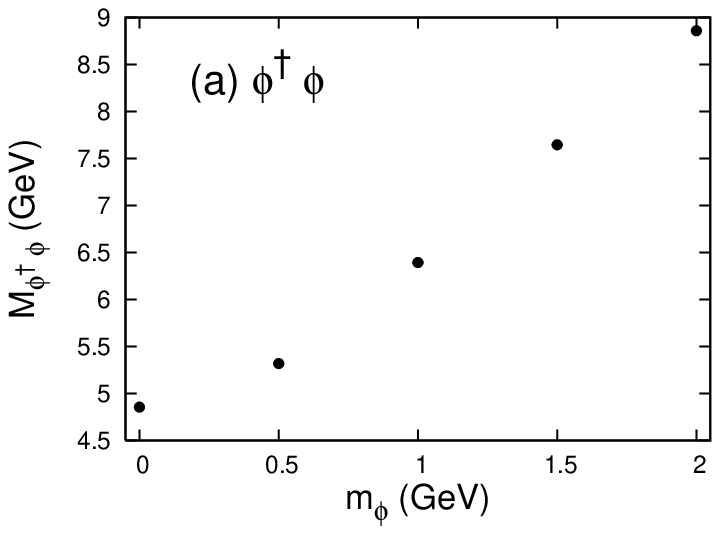} &
\includegraphics[width=5.5cm]{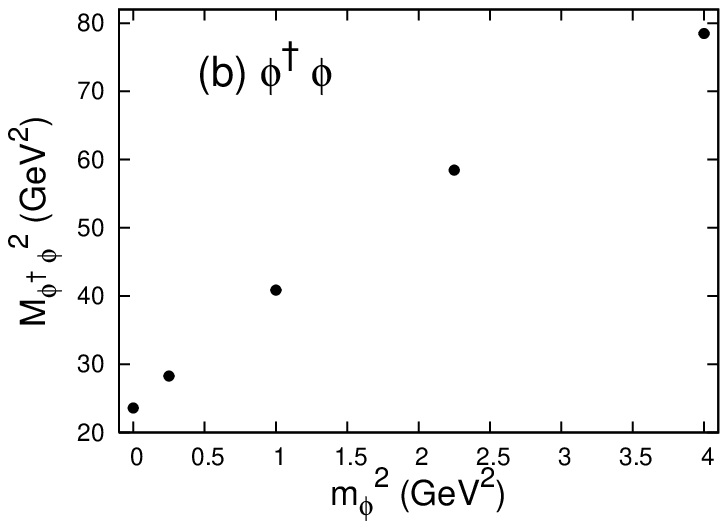} 
\end{tabular}
\caption{
The lattice results for scalar-quark mesons $\phi^\dagger \phi$ 
at $\beta=6.10$: 
(a) The scalar-quark meson mass $M_{\phi^\dagger\phi}$ 
plotted against the bare scalar-quark mass $m_{\phi}$ and 
(b) $M_{\phi^\dagger\phi}^2$ against $m_{\phi}^2$. 
A large mass of scalar-quark mesons 
about 5.5GeV even at $m_\phi=0$ and 
approximate linear $m_{\phi}^2$-dependence of 
$M_{\phi^\dagger\phi}^2$ are observed.
}
\label{fig12}
\end{center}
\end{figure*}

The lattice result for the scalar-quark mass at $\beta=6.10$ 
is summarized in Table VIII.
We show in Fig.13 the relation between 
the scalar-quark mass $M_{\phi^\dagger\phi}$ 
and the bare scalar-quark mass $m_\phi$ at $\beta=6.10$. 
The $m_\phi^2$-dependence of $M_{\phi^\dagger\phi}^2$ 
is found to be approximately linear.
Similar to the case at $\beta=5.70$, 
this behavior can be explained with 
the constituent scalar-quark picture of 
$M_{\phi^\dagger \phi} \simeq 2M_\phi$ 
and 
$M_\phi^2 \simeq m_\phi^2+\Sigma_\phi$, 
with $\Sigma_\phi$ being almost $m_\phi$-independent.
At $\beta$=6.10, the scalar-quark meson mass  
$M_{\phi^\dagger\phi}$ at $m_\phi=0$ 
is found to be about 5.5GeV, 
and thus the self-energy for scalar-quarks is 
estimated as $\sqrt{\Sigma_\phi}\simeq 2.8{\rm GeV}$, 
which is about twice larger than that at $\beta$=5.70. 

We compare the lattice result at $\beta=6.10$ with that at $\beta=5.70$ 
for the scalar-quark meson mass $M_{\phi^\dagger \phi}$ in the case of 
$m_\phi=0$, because of $M_{\phi^\dagger \phi} \simeq 2\sqrt{\Sigma_\phi}$ 
in the constituent picture.
For $m_\phi=0$, we find the approximate scaling relation as 
\begin{eqnarray}
\frac{(M_{\phi^\dagger\phi}a)|_{\beta=6.10}}
{(M_{\phi^\dagger\phi}a)|_{\beta=5.70}} \simeq 0.9,
\end{eqnarray}
which indicates a large quantum correction as 
$\sqrt{\Sigma_\phi} \propto a^{-1}$ mentioned in Sec.~\ref{sec6}. 
In this way, if one considers the point-like scalar-quark, 
its self-energy $\Sigma_\phi$ diverges in the continuum limit, 
and hence, to eliminate the divergence depending on the cutoff, 
one needs a fine tuning of the bare scalar-quark mass $m_\phi$.
For the argument of the diquark, however, there appears a natural cutoff 
corresponding to its intrinsic size as a composite object.


\begin{thebibliography}{99}
\bibitem{H64}
P.~W.~Higgs, Phys. Lett. {\bf 12}, 132, (1964).

\bibitem{PDG}
Particle Data Group (W.M.~Yao et al.), J. Phys. G{\bf 33}, 1 (2006).

\bibitem{RGG75}
A. De Rujula, H. Georgi, and S. L. Glashow, 
Phys. Rev. D{\bf 12}, 147(1975).

\bibitem{NJ61}
Y.~Nambu and G.~Jona-Lasinio, Phys. Rev. {\bf 122}, 345 (1961); 
Phys. Rev. {\bf 124}, 246 (1961).

\bibitem{H84}
K.~Higashijima, Phys. Rev. D{\bf 29}, 1228 (1984); 
Prog. Theor. Phys. Suppl. {\bf 104}, 1 (1991).

\bibitem{M83}
V.A.~Miransky, Sov. J. Nucl. Phys. {\bf 38}(2), 280 (1983); 
``Dynamical Symmetry Breaking in Quantum Field Theories'', 
World Scientific (1993).

\bibitem{KSWGSSS83}
J.B.~Kogut, M.~Stone, H.W.~Wyld, W.R.~Gibbs, J.~Shigemitsu, 
S.H.~Shenker, and D.K.Sinclair, Phys. Rev. Lett. {\bf 50}, 393 (1983).

\bibitem{HK94}
T.~Hatsuda and T.~Kunihiro, Phys. Rept. {\bf 247}, 221-367 (1994).

\bibitem{IOS05}
H.~Iida, M.~Oka, and H.~Suganuma, Eur. Phys. J. A{\bf 23}, 
305 (2005); Nucl. Phys. {\bf B141} (Proc. Suppl), 191 (2005).

\bibitem{DK87}
C.~DeTar and J.B.~Kogut, Phys. Rev. Lett. {\bf 59}, 399 (1987);
 Phys. Rev. D{\bf 36}, 2828 (1987). 

\bibitem{HL92}
T.~Hatsuda and S.H.~Lee, 
Phys. Rev. C{\bf 46}, R34 (1992).

\bibitem{CERES}
CERES Collaboration (G.~Agakichiev et al.), 
Phys. Rev. Lett. {\bf 75}, 1272 (1995).

\bibitem{KEK}
R.~Muto et al., Nucl. Phys. {\bf A774}, 723 (2006). 

\bibitem{MO87}
J.E.~Mandula and M.~Ogilvie, Phys. Lett. {\bf B185}, 127 (1987). 

\bibitem{AS99}
K.~Amemiya and H.~Suganuma, Phys. Rev. D{\bf 60}, 114509 (1999). 

\bibitem{MP99ISM02}
C.J.~Morningstar and M.~Peardon, 
Phys. Rev. D{\bf 60}, 034509 (1999).

\bibitem{ISM02}
N.~Ishii, H.~Suganuma, and H.~Matsufuru, 
Phys. Rev. D{\bf 66}, 094506 (2002); 
Phys. Rev. D{\bf 66}, 014507 (2002).

\bibitem{CR81}
F.E.~Close and R.G.~Roberts, Z. Phys. C{\bf 8}, 57 (1981). 

\bibitem{FJL82}
S.~Fredriksson, M.~Jandel, and T.~Larsson, Z. Phys. C{\bf 14}, 
35 (1982).

\bibitem{P76}
M.I.~Pavkovic, Phys. Rev. D{\bf 14}, 3186 (1976). 

\bibitem{W06}
F.~Wilczek, ``Diquarks as Inspiration and as Objects'', 
hep-ph/0409168.

\bibitem{BL84}
D.~Bailin and A.~Love, Phys. Rept. {\bf 107}, 325-385 (1984).

\bibitem{ARW99}
M.~Alford, K.~Rajagopal, and F.~Wilczek, Nucl. Phys. 
{\bf B537}, 443 (1999).

\bibitem{CL}
T.P.~Cheng and L.F.~Li, ``Gauge theory of elementary particle 
physics'', 
Oxford University Press (1988). 

\bibitem{IST06}
H.~Iida, H.~Suganuma, and T.T.~Takahashi, hep-lat/0612019, 
AIP Conf. Proc. (2007) in press.

\bibitem{TS0102}
T.T.~Takahashi, H.~Matsufuru, Y.~Nemoto, and H.~Suganuma, 
Phys. Rev. Lett. {\bf 86}, 18 (2001); 
T.T.~Takahashi, H.~Suganuma, Y.~Nemoto,  and H.~Matsufuru, 
Phys. Rev. D{\bf 65}, 114509 (2002);
H.~Suganuma, H.~Ichie and T.T.~Takahashi, 
Int. Conf. on ``Color Confinement and Hadrons in Quantum Chromodynamics", 249 
(World Scientific, 2004).

\bibitem{OST05}
F.~Okiharu, H.~Suganuma and T.T.~Takahashi, 
Phys. Rev. D{\bf 72}, 014505 (2005); Phys. Rev. Lett. {\bf 94}, 192001 (2005).

\bibitem{HN65}
M.Y.~Han and Y.~Nambu, Phys. Rev. {\bf 139}, B1006 (1965). 

\bibitem{Rothe}
H.J.~Rothe, ``Lattice gauge theories (3rd edition)'', 
World Scientific (2005), and references therein. 

\bibitem{EKM97}
A.X.~El-Khadra, A.S.~Kronfeld and P.B.~Mackenzie, 
Phys. Rev. D{\bf 55}, 3933 (1997).

\bibitem{Iida0605} 
H.~Iida, T.~Doi, N.~Ishii, H.~Suganuma and K.~Tsumura,
Phys. Rev. {\bf D74}, 074502 (2006);
N.~Ishii, T.~Doi, H.~Iida, M.~Oka, F.~Okiharu and H.~Suganuma, 
Phys. Rev. D{\bf 71}, 034001 (2005).

\bibitem{BCSVW94}
F.~Butler, H.~Chen, J.~Sexton, A.~Vaccarino, and 
D.~Weingarten, Nucl. Phys. {\bf B430}, 179 (1994).

\bibitem{TUOK05}
T.T.~Takahashi, T.~Umeda, T.~Onogi, and T.~Kunihiro, 
Phys. Rev. D{\bf 71}, 114509 (2005).

\bibitem{hierarchy}
M.E.~Peskin and D.V.~Schroeder, 
``An Introduction to Quantum Field Theory'', 
Perseus Books Publishing (1995).

\bibitem{DIOS04}
T.~Doi, N.~Ishii, M.~Oka and H.Suganuma, Phys. Rev. D{\bf70}, 034510 (2004).

\end{thebibliography}
\end{document}